\documentclass[aps,prx,preprint,groupedaddress]{revtex4-2}
\usepackage[top=2.5cm,bottom=2.5cm,right=3cm,left=3cm]{geometry}
\usepackage[utf8]{inputenc}
\usepackage{braket}
\usepackage{amsmath}
\usepackage{bbm}
\usepackage{appendix}
\usepackage{amsfonts}
\usepackage{comment}
\usepackage{cancel}
\usepackage{bbold}
\usepackage{graphicx}
\usepackage{float}
\usepackage{MnSymbol}
\usepackage{mathtools}
\usepackage{soul}
\usepackage{tabularx}
\usepackage{array}
\usepackage{accents}
\usepackage{bm}
\usepackage{placeins}
\usepackage{mathrsfs}

\newcolumntype{Y}{>{\centering\arraybackslash}X}

\usepackage[colorlinks, linkcolor=blue]{hyperref}
\usepackage[nameinlink, capitalise]{cleveref}

\newcounter{definition}
\newtheorem{Definition}[definition]{Definition}

\newcounter{theorem}
\newtheorem{Theorem}[theorem]{Theorem}

\newcounter{proposition}
\newtheorem{Proposition}[proposition]{Proposition}

\newcounter{lemma}
\newtheorem{Lemma}[lemma]{Lemma}

\newcounter{conjecture}
\newcommand{\domark}{%
	\vbox to 0pt{
		\kern-\dp\strutbox
		\hbox{\smash{\llap{*\kern1em}}}
		\vss
	}%
}

\begin{document}

	\title{Existence, structure, and properties of quantum-like states}
	\author{Gregory D. Scholes}
	\address{Department of Chemistry, Princeton University, Princeton, NJ 08544, USA}
	\email{gscholes@princeton.edu}

	\date{\today}

	\begin{abstract}
	    The main purpose of thispaper is to show that composite quantum-like (QL) systems can closely mimic the separable states of quantum systems, and that suitable physical systems exhibiting these states exist.  It is shown that QL graphs can closely emulate states of composite quantum systems, such as coupled two-level systems that display separable linear combinations of states. Examples of classical systems are suggested that show these states. These include multipole moments of waves or networks of phase oscillators. The work indicates that composite QL states can be manifest in complex network structures relevant to quantum biology or engineered into circuits, or even possibly soft matter. 
	\end{abstract}

	\maketitle
    \newpage


\section{Introduction}
Classical systems are ubiquitous and show remarkable diversity. They are the basis for almost all technologies and biological systems owing to their robustness and range of functions. Quantum systems, on the other hand, display special correlations that allow new functions\cite{Horodecki2009, NielsenChuang}. While classical systems can have some properties similar to those of quantum systems, such as wave-like interference and discrete spectra associated with eigenmodes, classical and quantum systems are fundamentally different. Here we study classical systems that have states similar to (`mimic', defined below) the \emph{states} of quantum systems. These entities are referred to as quantum-like (QL). How closely can a QL system emulate quantum states? The question is relevant for devising new kinds of technology, intermediate between quantum and classical, for computing and sensing. It is also pertinent to the search for quantum-like phenomena underlying biological functions in the field of quantum biology\cite{PlenioQBiol, Whatis}. Here the relevant question is how a composite QL state can be manifest in complex network structures or in soft matter. The main purpose of the present paper is to show that composite QL systems can closely mimic the separable states of quantum systems, and that suitable physical systems exist that exhibit these states.

An approach is to study a framework that allows systematic close comparison between classical and quantum composite states. In recent work, we have developed a suitable framework that allows classical systems to be designed that show the kinds of separable states needed for QL theories\cite{ScholesQLstates, QLproducts, QLsync, Debadrita1}. We have also studied how those systems might augment quantum information technology, albeit at the expense of resource size\cite{Amati1, Amati2}. The present work exploits the key idea from that work, that graph theory enables a representation for the physical structure of classical systems and a connection to associated states. Simply stated: graphs are physically realizable and have a spectrum. The process we use is to exhibit a graph that serves as an effective two-state system (a QL bit). Composite graphs and their states are generated systematically from this starting point to produce structures of classical systems that have states like those of composite quantum systems. Without the framework from graph theory, it would be difficult to search systematically for existence physical systems with the required states.  

Here we focus on the origin and properties of QL states. Prior work has explored other ways that classical systems can resemble their quantum counterparts. For example, how classical systems might exhibit probability laws similar to those of quantum systems. Khrennikov initiated an extensive program along these lines, exploring how and where QL probability laws can be found in diverse fields that include cognition, psychology, and finance\cite{Khrennikov2003, Khrennikov2005, Khrennikov2005b, Khrennikov1, Ozawa2020, KB2013, Khrennikov2016, Khrennikov2018SciRep, Khrennikov2023book}. At the core of that work is the identification of a kind of interference effect  can arise in probabilistic systems because of the way context changes between measurements. Other relevant work includes the cellular automata models studied by Elze\cite{Elze2017, Elze2024}, which can show intriguing correspondences to quantum mechanical systems. Further studies have shown how the interaction between a liquid droplet and waves in a liquid can lead to diffraction and interference that is produced in a strikingly similar way to diffraction and interference phenomena arising from a succession of single quantum particles\cite{Fort2006}. While these studies show how classical systems and classical theories can emulate aspects of quantum theory for functional gain, the present work is concerned with the question of how close the states of classical systems can get to analogous states of quantum composite systems.

\section{Definitions and Background}

\begin{Definition}{(\textcolor{green}{Quantum systems})}
	The class of quantum systems considered here are two-level (two-state) systems, analogous to quantum bits or ideal two level atoms. A quantum system labeled $i$ has states $v \in \mathbb{C}^2$ in the Hilbert space $\mathcal{H}_i$. 
\end{Definition}

\begin{Definition}{(\textcolor{green}{Composite state})}
	A finite number $q$ of quantum systems combine, possibly as a result of some kind of interaction, to produce composite states in the Hilbert space $\mathcal{H} = \mathcal{H}_1 \otimes \dots \otimes \mathcal{H}_q$ with dimension $2^q$.
\end{Definition}

\begin{Definition}{(\textcolor{green}{Quantum mimic or quantum-like system})}
	A classical system that we refer to as `quantum-like' (QL) has states that \emph{mimic} those of an analogous quantum system. That is, a QL bit labeled $i$ has states $v \in \mathbb{C}^2$ in the Hilbert space $\mathcal{H}_i$. The states represent elements of the group $\mathsf{SU}(2)$. See the proof of Prop. 1. Furthermore, a finite number $q$ of such QL bits can be combined  to produce composite states in the Hilbert space $\mathcal{H} = \mathcal{H}_1 \otimes \dots \otimes \mathcal{H}_q$ with dimension $2^q$. The QL classical system is constructed so that its spectrum emulates separable quantum \emph{states}, but it does not emulate the quantum system itself. 
\end{Definition}

A graph $G(n,m)$, that we often write simply as $G$, comprises $n$ vertices and a set of $m$ edges that connect pairs of vertices. The size of a graph or subgraph, that is, the number of vertices, is written $|G|$. The spectrum of a graph $G$ is defined as the spectrum (i.e. eigenvalues in the case of a finite graph) of the adjacency matrix $A$ associated with the graph. The spectrum of a graph is obtained by diagonalizing the corresponding adjacency matrix. The adjacency matrix rows and columns are indexed by the graph vertices. The diagonal entries are zero, while off-diagonal entries contain 1 at $a_{ij}$ if the vertex $i$ is linked by an edge to vertex $j$. Here we will specialize to undirected graphs, so the adjacency matrix is symmetric ($a_{ij} = a^*_{ji}$), but it is also possible to work with directed graphs. For background see ref \cite{Excitonics2024}. Below this this basic model is extended so that the edges may take any value on the unit circle in the complex plane\cite{Zaslavsky1, Mehatari, Reff2012}. We refer to this edge value in the adjacency matrix as the edge bias.

In the QL graph model, $d$-regular graphs are used:
\begin{Definition}
	($d$-regular graph) A graph $G$ is $d$-regular if every vertex has degree $d$. That is, every vertex connects to $d$ edges.
\end{Definition}

Furthermore, we will exploit the concept of expander graphs, which are guaranteed to have an emergent state, separated from all other states in the spectrum. An emergent state is a state whose eigenvalue is distinguished from the rest of the spectrum. It is often associated with collective properties or modes, as in the example of Ref. \cite{Kippenberg2024}. Expander graphs\cite{Sarnak2004, expandersguide, Lubotzky, Expanders, Expanders2, Alon1986, Tao-expanders} are highly connected graphs that are optimal for random walks and communications networks. The physical concept underpinning an expander is that the edges are scale-free, so that no matter how we lay out the vertices, edges connect vertices at all length scales.  $d$-regular \emph{random} graphs are likely to be expander graphs.

\section{Outline}

How can quantum-like states be produced by a classical system? We approach this question by proving the existence of classical systems with separable composite QL states. We have shown in prior work that composite QL states comprising arbitrary numbers of QL bits can be generated, in principle. In this paper, we show that the separable state vectors generated by the QL bits are defined in the Hilbert space $\mathcal{H}$ $\cong \mathbb{C}^2$, and can be rotated by continuous operation on the classical system. In terms of the QL graphs, that means continuously changing the edge biases. We further discuss what kinds of classical systems are needed to produce these states. This model is extended to composite states defined in $\mathcal{H}_1 \otimes \mathcal{H}_2 \otimes \dots$.

We identify examples of classical systems that have projected states that are close classical analogs to the quantum states of a two-level system. In particular, we propose classical systems with states that represent elements of the group $\mathsf{SU}(2)$. Then it is shown how such systems---QL bits---can be represented by graphs. The graph has structure and topology which can serve as a blueprint for any suitable classical system. Essentially, the graph defines how phases, correlations or couplings in the classical system need to be organized. This step allows the concept of a QL bit identified in any known classical system to be generalized to new kinds of classical systems or constructions exhibiting \emph{composite} QL states. A further advantage of a graph is that it is naturally associated with a matrix (we use the adjacency matrix). Hence, the graph connects a classical system it designs to a linearized state space. 

To ensure the states are provided in the tensor product basis, we build the composite classical systems systematically based on the Cartesian product of graphs, denoted $G_A \Box G_B \Box \dots$. That systematic construction shows how suitable classical systems can replicate all separable quantum states. The QL graph model for composite QL states, therefore, is a physical analog to the tensor product of quantum states.

\section{Physical Basis}

To motivate the development of a proof with a physical basis, it is shown that classical systems exist whose states represent elements of the group $\mathsf{SU}(2)$. The result is already known, but it is a useful foundation for what follows because it provides concrete examples of relevant physical systems. It is then proved that these systems can be represented much more generally as a particular graph, which ultimately allows generalization of the system to composite classical systems that display states analogous to corresponding composite quantum systems. In other work\cite{QLsync} it has been established how linearity of the states is obtained as a limit of strong-synchronization among the constituent phases of the classical system (which otherwise might be a nonlinear system). 

\begin{Proposition}
	Polarization states of classical waves are represented by the elements of the group $\mathsf{SU}(2)$.
\end{Proposition}

The polarization of transverse waves specifies the projection of the wave's amplitude perpendicular to the direction of motion of the wave. Given a basis of unit vectors, $|x\rangle, |y \rangle$, in the plane normal to the direction of motion, the fundamental states of polarization are well-known to be linear polarizations $|x\rangle$ and $|y \rangle$ and circular polarizations $(|x\rangle \pm i |y\rangle)/\sqrt{2}$. Intermediate (elliptical) polarizations are obtained as linear combinations of these polarization states. Examples of waves that exhibit such polarization states include light, radio waves, and elastic waves in matter. A recent example of the latter is generation of circularly-polarized elastic waves in an isotropic solid medium\cite{Elastic}. 

The polarization state of waves can be determined using a sequence of projective measurements to give the so-called Stokes parameters\cite{Kliger}. The polarization state of a wave is commonly described by the Jones vector,
\begin{align}
	J &= 	
	\begin{pmatrix}
		A_x e^{i \phi_x} \\
		A_y e^{i \phi_y}  
	\end{pmatrix} \\
	&= 
	\begin{pmatrix}
		a + bi \\
		c + di  
	\end{pmatrix}	.
\end{align}
where $A_x$ and $A_y$ are amplitudes, $a, b, c, d \in \mathbb{R}$, and $\phi_x$ ($\phi_y$) is the angle of the vector in the complex plane. A 2-vector such as $J$, normalized so that $|J| = 1$, with entries in $\mathbb{C}$ can be mapped to a quaternion $q$ given by
\begin{align}
	q &= (a + bi) + (c +di)j \\
	&= a + bi + cj + dk,
\end{align}
normalized so that $|q| = a^2 + b^2 + c^2 + d^2 = 1$. It is well known\cite{NaiveLie, Cohn, Hall2015} that these unit quaternions form the group of rotations over $\mathbb{S}^3$ also known as the group $\mathsf{SU}(2)$. We can take $1 \rightarrow \sigma_0$, $i \rightarrow -\sigma_1$, $j \rightarrow -\sigma_2$, and $k \rightarrow -\sigma_3$, where $\sigma_i$ are the Pauli matrices, to give an isomorphism such that unit quaternions are viewed as the group of $2 \times 2$ complex unitary matrices $Q$:
\begin{equation}
	Q = 
	\begin{pmatrix}
		a + di & -b - ci \\
		b - ci & a -di  
	\end{pmatrix} ,
\end{equation}
where $\det(Q) = 1$ (see e.g. Ch. 6 of \cite{Woit}). Endowed with the operation of quaternion multiplication, $q$ are elements of the group $\mathsf{SU}(2)$.  

\section{Existence of classical quantum mimics with separable states}

\begin{Theorem}
	There exists classical systems that are associated with a measurable state space that mimics the \emph{separable} states of an analogous quantum system. Specifically, there exists classical systems incorporating a notion of phase relationships that can be transformed continuously such that the states of a composite quantum-like system represent elements of the group $\mathsf{SU}(2) \times \dots \times \mathsf{SU}(2)$.
\end{Theorem}

The first step of the proof is to show that the kinds of classical systems described in the previous section have a graph representation. This allows the concept of classical systems that have states that, when mapped to matrices in the obvious way, comprise elements of the group $\mathsf{SU}(2)$ to be generalized. That is, the graph gives a structure map class that can be specialized to various specific structures. For example, as we discuss later in the paper, the classical system could be a tensor multipole moment, a network of phase oscillators, an elastic medium, and so on. But, in each case it can be represented by same graph.

In particular, what we need to do is find a map from a classical system to a state space defined in a tensor product basis and that can accommodate (classical) superpositions. 
  
\begin{Lemma}
 	The polarization states of classical systems and their continuous transformations as elements of the group $\mathsf{SU}(2)$ are represented by a class of graphs that we call QL bit graphs.
\end{Lemma}

\begin{figure}
	\includegraphics[width=6.5 cm]{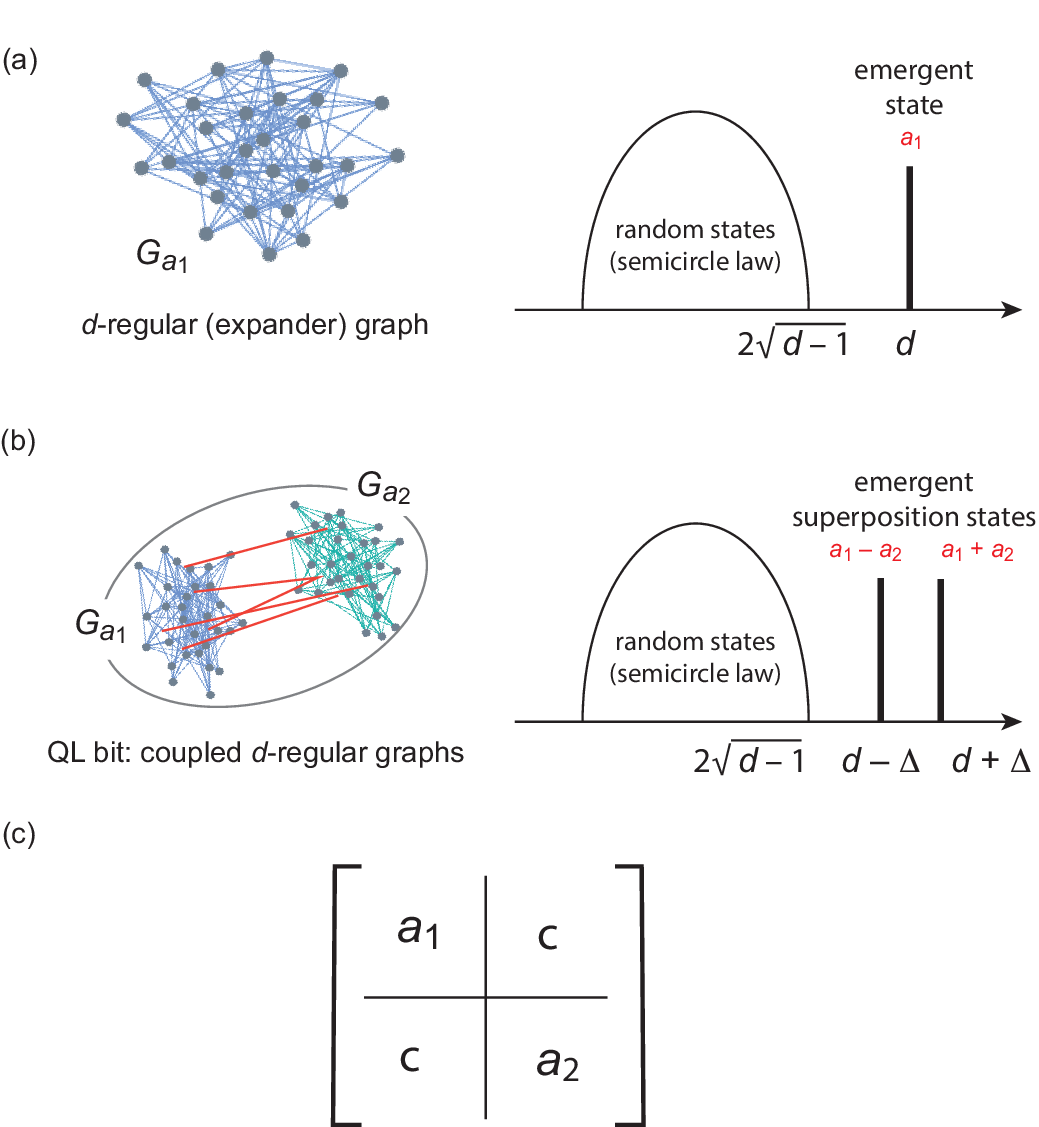}
	\caption{(a) Drawing of a small $d$-regular graph and the spectrum representative of a large $d$-regular graph, showing that the single emergent state is separated in the spectrum from the many other states that we refer to as `random states'. (b) A QL bit is constructed by coupling together two $d$-regular subgraphs. The coupling edges, shown in red, are added randomly from each vertex in $G_{a1}$ to each vertex in $G_{a2}$ with probability 0.2. Realistically, the QL bit will likely not show the subgraphs separated in space, like we display here for clarity; instead the vertices can be positioned randomly. (c) Adjacency matrix of a QL bit showing the diagonal blocks hosting the adjacency matrices for each subgraph. These blocks are coupled by edges in the off-diagonal blocks labeled $c$ that hybridize the subgraphs.}
	\label{figColloq2}
\end{figure}

We define a QL bit graph as follows\cite{ScholesQLstates}. Combine two expander graphs by coupling them using new edges, Fig. \ref{figColloq2}. These coupling edges are the red edges drawn in Fig. \ref{figColloq2}b denoted collectively as the blocks $\mathbf{c}$ in the adjacency matrix. As long as we include a small number of coupling edges compared to the edge density in each of these subgraphs, we can thereby hybridize the emergent states, producing two new emergent states that are the in- and out-of-phase linear combinations of the emergent states for each subgraph in isolation. 

The way to understand this graph construction is to think about its adjacency matrix. The adjacency matrix has a block form, shown in Fig. \ref{figColloq2}c. Each of the blocks $[\mathbf{a}_1]$ and $[\mathbf{a}_2]$ is an adjacency matrix for an expander graph, so each one of the these blocks has a distinguished emergent state, like shown in Fig. \ref{figColloq2}a. By connecting these blocks with the off-diagonal `coupling' blocks $[\mathbf{c}]$, these two emergent states are coupled. The emergent states of the QL bit are therefore the two linear combinations of these states, Fig. \ref{figColloq2}b. By defining a basis in this block form, we obtain the effective states of a two-level system:
\begin{equation}
	W_{2 \times 2} = \alpha |a_1 \rangle + \beta |a_2 \rangle .
\end{equation}

The simplest definition of a QL bit graph considers all the edges to be given by a value of 1 in the adjacency matrix. That standard assignment is easily generalized. For example, we can define a graph as follows.

\begin{Definition}
	A complex unit gain graph\cite{Reff2012, Mehatari} is a graph where each edge is assigned a complex unit. That is, the edge from vertex $i$ to vertex $j$ takes a value $e_{ij} \in \mathbb{T}$, where $\mathbb{T} = \{ z \in \mathbb{C}: |z| = 1 \}$. The edge in the opposite direction is the inverse, that is, $e_{ij} = e^{-1}_{ji} = \bar{e}_{ij}$, where the bar indicates complex conjugate. The (Hermitian) adjacency matrix of the complex unit gain graph therefore contains entries in the upper diagonal $e_{ij} \in \mathbb{T}$, and corresponding complex conjugates in the lower diagonal. 
\end{Definition}

In order to generate states of the QL bit that have complex coefficients, we need to use edges defined for the complex unit gain graph for the connecting edges. We allow connecting edges (i.e. the blocks $[\mathbf{c}]$) to take values from the unit circle in the complex plane, $\mathbb{T} = \{ z \in \mathbb{C}: |z| = 1 \}$. Then, continuous rotation of the connecting edge bias generates states that represent the elements of $\mathsf{SU}(2)$, as the vectors $(\alpha, \beta)$, where $\alpha$ and $\beta$ are the coefficients of $W_{2 \times 2}$. The corresponding elements of  $\mathsf{SU}(2)$ are the matrices $\begin{psmallmatrix} \alpha & -\bar{\beta}\\ \beta & \bar{\alpha}\end{psmallmatrix}$. That is demonstrated as follows.

The QL bit graph $G_A$ can be generalized to a block form of an effective $2 \times2$ $d$-regular random unit gain graph by structuring its adjacency matrix as follows:
\begin{equation}
	A_A = 
	\begin{pmatrix}
		\mathbf{a}_1 & \mathbf{c}_{12} \\
		\mathbf{c}_{21} & \mathbf{a}_2  
	\end{pmatrix} ,
\end{equation}
where blocks $\mathbf{a}_1$ and $\mathbf{a}_2$ define the $d$-regular random subgraphs with edge bias = 1. The edges in block $\mathbf{c}_{12}$ all have identical bias, taking some value $c \in \mathbb{T}$, while the edges in block $\mathbf{c}_{21}$ take the value $\bar{c}$. The eigenvectors (but not the eigenvalues) of the corresponding $W_{2 \times 2}$ are equivalent to those of the matrix
\begin{equation}
	A_{2 \times 2} =
	\begin{pmatrix}
		0 & c \\
		\bar{c} & 0  
	\end{pmatrix}.
\end{equation}
This simple representation is possible because each of the sub-blocks $\mathbf{a}_1$ and $\mathbf{a}_2$ correspond to an adjacency matrix of a $d$-regular graph and hence produce a single emergent state with eigenvalue $d$. The off-diagonal blocks $\mathbf{c}_{12}$ and $\mathbf{c}_{21}$ couple these graphs symmetrically, giving an emergent state that is a linear combination of the emergent states of the $\mathbf{a}$-blocks. Then it is readily observed that the effective $2 \times 2$ emergent states are given by Eq. 8.

It is seen that, up to an overall arbitrary phase and neglecting the obvious normalization factor, the state with highest eigenvalue (eigenvalue = 1) of $A_{2 \times 2}$ is $v = |a_1\rangle \pm |a_2\rangle$ when $c = \pm 1$, or $v= |a_1\rangle \pm i|a_2\rangle$ when $c = \mp i$. Obviously these states are analogous to those of linearly- or circularly-polarized waves. More generally, for any $c \in \mathbb{T}$, the emergent state, up to an arbitrary overall phase, of a QL bit is $v = (|a_1\rangle + \bar{c}|a_2\rangle)/\sqrt{2}$. Hence there is a one-to-one map from the connecting edge bias to the eigenstate coefficients that generates all elements of the group $\mathsf{SU}(2)$. 

The QL bit graph represents systems analogous to polarized classical waves. It can thus reveal other classical systems with the same kinds of states. For example, the QL bit graph representation provides a basis for constructing representations of classical systems with more complicated polarization states, analogous to multipole moment tensors. In particular, we can combine $q$ QL bit graphs using the Cartesian product of graphs to produce the structure of classical systems with separable states in $\mathcal{H} = \mathcal{H}_1 \otimes \mathcal{H}_2 \otimes \dots \otimes \mathcal{H}_q$, as we previously reported\cite{QLproducts}. Thus,

\begin{Lemma}
	A classical system represented by the Cartesian product of QL bit graphs (a composite QL system) has emergent states equivalent to the separable states of a composite quantum system. These states are the separable superpositions in an appropriate tensor product basis.
\end{Lemma}

The Lemma is proved, after defining the Cartesian product of graphs\cite{GraphProducts,Sabidussi} (see also Appendix B of Ref. \cite{QLsync}), by a known Proposition that relates the states of a Cartesian product of two graphs ($G \Box H$) to the tensor product of the states of each of the graphs $G$ and $H$. 

\begin{Definition}
	(Cartesian product of graphs) $G \Box H$ is defined on the Cartesian product of vertex sets, $V(G) \times V(H)$. Let $\{u, v, \dots\} \in V(G)$ and $\{x, y \dots\} \in V(H)$. Let $E(G)$ and $E(H)$ be the set of edges in $G$ and $H$ respectively. The edge set of the product graph $G \Box H$ is defined with respect to all edges in $G$ and all edges in $H$  as follows. We have an edge in $G \Box H$ from vertex $(u,x)$ to vertex $(v,y)$ when
	\begin{itemize}
		\item either there is an edge from $u$ to $v$ in $G$ and  $x = y$,
		\item or there is an edge from $x$ to $y$ in $H$ and  $u = v$.
	\end{itemize}
\end{Definition}

A basis for the $2^q$  states in $\mathcal{H}$ that is produced by the Cartesian product of QL bits is 
\begin{equation}
	|a_i\rangle \otimes |b_j\rangle \otimes |c_k\rangle \otimes \dots ,
\end{equation}
where $i, j, k, \dots \in \{1, 2\}$ and states $a_i$ come from subgraphs of graph $G_A$, states $b_j$ come from subgraphs of graph $G_B$, and so on. This result is evident from the following proposition.

\begin{Proposition}\label{eq:eig_prod}
	(Spectrum of a Cartesian product of graphs) Given
	\begin{enumerate}
		\item[] A graph $G$, for which its adjacency matrix $A_G$ has eigenvalues $\lambda_i$ and eigenvectors $X_i$, and
		\item[] A graph $H$, for which its adjacency matrix $A_H$ has eigenvalues $\mu_i$ and eigenvectors $Y_i$, then
	\end{enumerate}
	the spectrum of $G \Box H$ contains eigenvalues $\lambda_i + \mu_j$ and the corresponding eigenvectors are $X_i \otimes Y_j$.
\end{Proposition}

The key result is that, given an eigenvector, say $v_A$, for the emergent state of one QL bit with graph $G_A$ and an eigenvector $v_B$ for the emergent state of another QL bit with graph $G_B$, then we have an eigenvector $v_A \otimes v_B$ for the graph product $G_A \Box G_B$. Hence we have a map between a graph---the product graph---and a tensor product in the state space. Now, since the emergent states of each QL bit represent elements of $\mathsf{SU}(2)$, then the states of the QL bit Cartesian graph product represent elements of $\mathsf{SU}(2) \times \dots \times \mathsf{SU}(2)$. These are the separable states of the corresponding composite quantum system, as required. That is, the graph identifies the structure of classical systems whose states mimic those of the subset of separable states of analogous quantum systems such as coupled two-level systems. The \emph{structure} refers to the way connecting edges in the graph install phase relationships among components of the system. 

As a side-note, the graphs so far described all have unit magnitude weighted edges. However, we can choose to weight edges by any scalar. The adjacency matrix is analogous to a Hamiltonian matrix for the system, where we may assign values to the diagonals to indicate energies or frequencies of the basis states and couplings to the off-diagonal elements. The graphs relate physical structure to this Hamiltonian\cite{Scholes2020} and, importantly, they define phase relationships between the basis states. The QL bit graphs can thus be produced by multiplying the edges in the QL bit adjacency matrix by a scalar before taking the product. A positive factor $<1$ shifts the emergent eigenvalue and therefore `de-tunes' the resonance of this graph with others, and reduces its weight in the emergent state of the Cartesian product.

To summarize, the QL graph representation of classical systems was used to refine the search for classical systems that display states that mimic those of quantum systems. It has thus been proven that, by tuning the connecting edge bias in a QL bit graph, a continuous set of states is generated that represent all elements of the group $\mathsf{SU}(2)$. It was also demonstrated that representative classical systems are already well-known. 

The graph representation allows classical systems to be identified that have composite states that mimic those of  \emph{composite} quantum systems. In particular, it was proved that the Cartesian product of $q$ QL bit graphs generates graphs whose pure states represent elements of the group $\mathsf{SU}(2) \times \dots \times \mathsf{SU}(2)$ (the $q$-fold product). The graph can be directly interpreted as a classical network of coupled phase oscillators, where the complex coupling edge bias is defined by the phase difference between oscillators in different subgraph blocks \cite{QLsync}. Such systems are known to exist and can be demonstrated as electronic circuits\cite{topcircuit2, topcircuit1}.

This completes the proof of Theorem 1. 

\subsection{Examples of possible classical systems with composite QL states}

We have proved that there exist classical systems that have states that mimic the separable states of analogous composite quantum systems. Where and how would we look for these systems? Here we discuss possible manifestations of the QL graphs. 

A notion of phase is an important requirement for any QL classical system. Thus, waves and wavelike systems are obvious candidates for consideration. Waves are ubiquitous, and include light, radio waves, acoustic, and elastic waves in matter. As proposed above, relevant states of the waves are dipolar polarization states, which are isomorphic to the states of the QL bit. 

A challenge is to extend wave polarization to tensor products of polarization in physical systems. This step is needed in order to find analogies to QL \emph{composite} states.  The waves are confined to three dimensions, but the required basis is much larger, which suggests they must have states characterized by tensors. Physical examples of classical waves with multipolar polarization states include charge multipoles, acoustic multipoles, and strain multipoles. The latter are related to nonlinear elasticity\cite{nonlinelast} and have been studied extensively in emulsions comprising colloids or fluid droplets in a nematic liquid crystal host\cite{smalyukh1, smalyukh2}. Quadrupolar waves are also known in solid state condensed matter\cite{carretta}. Conceptually, we can associate a phase quadrupole with the product graph shown in Fig. \ref{figExistQuad}a. Notice that the phase differences are uncorrelated. That is, in the case of quadrupolar strain, the strain variation along one tensor index is independent of that along the other index. 

An alternative viable class of systems includes networks and circuits. The adjacency matrix of a graph is analogous to a Hamiltonian matrix for the system\cite{Scholes2020}. That is, we may assign values to the diagonal elements to indicate energies or frequencies of the basis states and couplings are defined by graph edges. The graphs give structure to this Hamiltonian by defining which basis states are pairwise coupled and, importantly, they define phase relationships between the basis states. With this picture in mind, networks of interacting systems, conveniently viewed as networks of phase oscillators\cite{QLsync, Strogatz2000} are one example of such classical systems\cite{QLsync}. Another example is an electronic circuit with complex edge bias. These kinds of circuits have been developed in the field of `topological circuits'\cite{topcircuit1, topcircuit2}. Optical analogs can also be envisioned, as discussed in Sec. VI. One can speculate whether various other complex networks or neural network structures could similarly be designed to produce QL states\cite{DecoQLbrain}. The circuit model also provides a template for the structure of more complex systems, potentially connecting the QL model to surprising systems such as oscillations in slime moulds\cite{slime1}.

To leverage QL correlations for a functional advantage, though, we must work out how to access them. Unlike a quantum mechanical system, we cannot generally decompose the quadrupole moment into two dipolar contributions (and so on for higher multipoles). Therefore, we cannot perform joint measurements in the local basis of each subsystem. These kinds of systems are  \emph{classical}.

\subsection{An example of properties of QL states}

As an example of how QL states have properties like the corresponding quantum states, let's consider the well-known no-cloning theorem\cite{nocloning}.

\begin{Theorem}
	(Wooters and Zurek) There is no unitary operator $U$ on $\mathcal{H} \otimes \mathcal{H}$ such that for all normalized states $|a \rangle_A$ and $|b\rangle_B$ in $\mathcal{H} = \mathcal{H}_A = \mathcal{H}_B$
	\begin{equation*}
		U\Big( |a\rangle_A |b\rangle_B \Big) = e^{i \alpha(a,b)} |a\rangle_A  |a\rangle_B
	\end{equation*}
	where $\alpha \in \mathbb{R}$.
\end{Theorem}

If $|a\rangle_A$ is a state of the QL graph $G_A$ and $|b\rangle_B$ is a state of the QL graph $G_B$, then $|a\rangle_A \otimes |b\rangle_B$ is a state of the Cartesian product of those graphs, $G_A \Box G_B$. Recall that each vertex of $G_A$ (alternatively $G_B$) is now the graph $G_B$ (alternatively $G_A$), with additional edges added appropriately (see Appendix B of Ref. \cite{QLsync}). For every vertex of graph $G_A$, we insert the adjacency matrix of $G_B$.  The adjacency matrix of  $G_A \Box G_B$ can be drawn in block form in such a way that the inherited edge structure from $G_B$ is highlighted on the diagonal:
\begin{equation}
	A_{A\Box B} = 	
	\begin{bmatrix}
		A_B & * & * & * & \dots \\
		* & A_B & * & * & \dots \\
		* & * & A_B & * & \dots \\
		* & * & * & A_B & \dots \\
		\vdots & \vdots & \vdots & \vdots & \ddots
	\end{bmatrix}
\end{equation}
and all the other blocks, written as $[*]$ contain edge structure inherited from $G_A$ or are zero. Let's index the blocks $[A_B]_{ii}$, $i = 1, \dots, n_0$, and assume that each $G_A$ and $G_B$ each contain $n_0$ vertices. A $[*]_{ij}$ block is non-zero if there is an edge in $G_A$ connecting vertex $i$ to vertex $j$.

When translated to graphs of QL states, the no-cloning theorem says that there is no unitary matrix that can transform $A_{A\Box B}$ such that, to an overall phase factor, diagonal blocks $[A_B]$ are mapped to copies of $[A_A]$, while leaving all the blocks $[*]$ unchanged.

Let's assume that $[A_A]$ and $[A_B]$ have the same dimension $n_0$ (i.e. the same number of vertices). Further, we will give each graph the same edge set to simplify the problem further, but the graphs are distinguished by the edge bias of each QL bit, which can be encoded as a phase $\theta_i$ assigned to each vertex\cite{QLsync}. The unitary operator we consider is then, 
\begin{equation}
	\Phi = 
	\begin{bmatrix}
		e^{i\theta_1} & 0 & 0 & \dots \\
		0 & e^{i\theta_2} & 0 & \dots \\
		0 & 0 & \ddots & 0 \\
		0 & \dots & 0 & e^{i\theta_{2n}} 
	\end{bmatrix}
\end{equation}

The transformation, $A_{A\Box B}^{\prime} = \Phi^{-1}A_{A\Box B}\Phi $, should change the edge bias in all the $[A_B]$ blocks (to turn them into $[A_A]$ blocks), without changing the $[*]$ blocks.  We just need to consider the effect of the map on a few elements of the matrix to understand why cloning the state $|a\rangle_A$ is impossible. The map $\Phi^{-1}A_{A\Box B}\Phi$ introduces a phase rotation $e^{i(\theta_2 - \theta_1)}$ in the first off-diagonal element of the top-left $[A_B]$ block, and a rotation $e^{i(\theta_n - \theta_m)}$, where $m = n_0+1$ and $n = n_0+2$, in the analogous element of the second $[A_B]$ block. These blocks must be changed identically by the map, so $\theta_2 - \theta_1 = \theta_n - \theta_m$, or equivalently $\theta_m - \theta_1 = \theta_n - \theta_2$.

Now let's consider the first $[*]$ block, which we require to be unchanged by the map because $|a\rangle_A$ should not be changed during the cloning except by a constant phase. The first two elements of the top row of $[*]$ are  $e^{i(\theta_m - \theta_1)}$  and $e^{i(\theta_n - \theta_1)}$. Thus we require that $\theta_m - \theta_1 = \theta_n - \theta_1$, that is, $\theta_m = \theta_n$. Clearly this us impossible to achieve if we want the $[A_B]$ blocks to be transformed. Therefore, even though the underlying network encoded by the graph is classical, there is no way of changing the phases of its constituent phase oscillators that allows a QL bit state to be cloned. 

This result illustrates how the structure of the adjacency matrix of a composite QL system is the important factor that produces effective states that are `quantum like'. The QL states and associated adjacency matrix satisfy Theorem 2. However, interestingly, the underlying graph is a classical object and it should be possible to clone it.

\section{Non-separable QL states}

It is well-known that classical systems can show non-separable states, for example, when interactions in the course of time prevent functions of two variables being written as a product. Non-separable states in classical systems have been studied, especially,  for  classical light that is suitably structured\cite{ForbesNatPhoton} to produce what has been termed \emph{classical entanglement}\cite{nonseplight, Forbes2019}. Basis states can be polarization, spatial mode, orbital angular momentum (a characteristic of `twisted light'\cite{Barnett-twisted}), etc. Work has shown that interference of suitably prepared beams\cite{Spreeuw1998} or light beams with spatially inhomogenous states of polarization\cite{Forbes2015, KonradForbes} can produce nonseparable states of light. The consensus view is that the states are locally-correlated (i.e. nonlocality is not present). According to Ref. \cite{Brunner} we expect to have local correlations if and only if the correlations are classical, meaning a Bell-type inequality is not violated. Thus, these non-separable states of a wave are entirely classical and therefore cannot exhibit uniquely quantum correlations such as entanglement. However, classically-entangled light has been shown to be a good mimic for quantum systems and potentially a related resource\cite{ForbesQChannels, ForbesQcomp}. In Ref. \cite{Forbes2019}, Forbes and co-workers summarize how to think about these states, and we quote here three of those points:

\begin{quote}
	\begin{itemize}
		\item Classical entanglement denotes a set of ideas which establish a connection between some properties of the representation of quantum systems and the same properties of the representation of classical systems.
		\item Classical entanglement does not lay at the border between quantum and classical physics, but it is entirely contained within the domain of classical physics.
		\item Classical entanglement is strictly local (viz., there is not something like ``classical nonlocality'').
	\end{itemize}
\end{quote}

Spreeuw exhibits a set-up that appears to violate the CHSH-type inequality, suggesting that the classical light is in an entangled state. A fundamental concern Spreeuw raises, however, is that one of the cebits (QL bits) cannot be separated from the other, which is the main issue with proposals for classical entanglement. That observation contrarily suggests that the correlation is fundamentally local and it should \emph{not} violate a Bell-type inequality\cite{Brunner}. That is, these are non-separable states, but \emph{not} entangled states. Paneru and co-workers\cite{Paneru2020} give a survey of entanglement versus `classical entanglement'. They argue that ``classical correlations cannot lead to the same conclusions as quantum entanglement''. They also note the importance of being able to separate spatially the qubits (QL bits in our case). Paneru and co-workers\cite{Paneru2020} make the important observation that quantum systems are comprised of particles, like photons or electrons and so on, that display wave-particle duality. This is indeed a crucial difference between quantum systems and QL systems generally. 

Here we ask how the QL graph model can be extended, for example by incorporating additional edges, so that non-separable states might be accessed. We first show that the adjacency matrices of the separable QL states have a particular structure. We find that adding edges to blocks forbidden for separable states gives non-separable states. It is argued that a physical basis for adding these kinds of correlations (but strictly speaking, not literally adding edges in that way) is to account for many-body correlations. This is shown by considering and extending the Kuramoto model for a network of phase oscillators. Finally, it is shown that maximally entangled states appear to be associated with disconnected graphs, emphasizing that classical mimics of entangled states likely cannot be obtained from classical graphs. 

The number of vertices in each QL bit graph can be any number $N$, and we will assign $N/2 = n$ of these vertices to each subgraph (although we are free to partition the graphs in other ways).  Composite QL graphs are represented by their $N^q \times N^q \times \dots$ adjacency matrices. These matrices have a block structure, where blocks containing all entries of 0 (zero-coupling blocks) correspond to edge blocks in the adjacency matrix indicating subgraphs in the Cartesian product that are not directly connected to each other. These are the evident as the blocks that would otherwise couple product states that differ by more than one index. For example, in the three-QL bit graph the product basis block $a_1b_1c_1$ does not couple directly to the block $a_2b_2c_1$. On the other hand, all coupling blocks contain a finite number of entries of $c \in \mathbb{T}$.

\begin{figure}
	\includegraphics[width=6.5 cm]{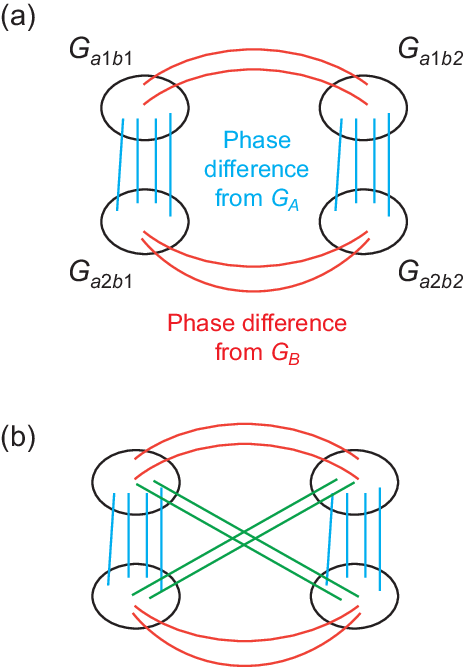}
	\caption{(a) Schematic representation of the graph formed by the Cartesian product of two QL bit graphs. (b) The same graph, but with edges added to indicate higher order correlations.}
	\label{figExistQuad}
\end{figure}

The 2-QL bit graph adjacency matrix contains two zero-coupling blocks above the diagonal and, correspondingly, two below the diagonal. The 3 QL bit graph contains 16 zero-coupling blocks above and 16 below the diagonal. By realizing that the QL graph for a composite system has the same structure as that of a Boolean poset (see Appendix A), then it is straightforward to see that the total number of zero-coupling blocks for a $q$-fold Cartesian product of QL bit graphs in either the upper or lower diagonal of the matrix equals the number of edges in a complete graph on $2^q$ vertices (the number of subgraphs in the $q$-fold Cartesian product of QL bits) minus the number of edges in a $q$-dimensional hypercube. That is, the number of zero-coupling blocks in the upper (or lower) diagonal part of the adjacency matrix is $\frac{1}{2}2^q(2^{q} - 1) - \frac{1}{2}q2^q$ = $\frac{1}{2}(2^{2q} - [q+1]2^q)$. 

Noting the definitions from previous sections, then for any $N$, these matrices just defined form a subspace defining all the separable pure emergent states with one-to-one correspondence to the elements of the group $\mathsf{SU}(2) \times \dots \times \mathsf{SU}(2)$. However, the \emph{total} space of QL states includes separable and non-separable states. For the kinds of QL state graphs that we have considered, those that exhibit non-separable states might be associated with adjacency matrices of separable QL graphs where  zero-coupling blocks become coupling blocks. A physical basis that conceptually `flattens' to qualitatively adding edges in that way is suggested by considering and extending the Kuramoto model for a network of phase oscillators as follows.

The Kuramoto model for phase oscillator synchronization is well-known and widely explored\cite{Strogatz1993, Strogatz2000, StrogatzBook, Acebron2005}. Each vertex of a QL graph can be thought of as an oscillator endowed with a frequency, that we treat in the rotating frame of the network of oscillators by its difference from the mean frequency $\epsilon_i$, and a phase offset $\phi_i$ for the oscillator at vertex $i$. These terms give the time-dependent phase for each oscillator $\theta_i = \epsilon_it + \phi_i$, which is associated with the set of vertices. The oscillators are coupled according to the edges in the graph which, under appropriate conditions\cite{Strogatz2000, Acebron2005, daFonseca2018, Rodrigues2016, ScholesAbsSync}, allows the oscillators to synchronize after several periods of time. The phases associated to the vertices come into play as phase differences $e^{i(\theta_i -\theta_j)}$ multiplying the non-zero entries off-diagonal of the graph's adjacency matrix. This is accomplished by a suitable unitary transformation\cite{QLsync}. The phases then evolve, according to the Kuramoto model\cite{daFonseca2018, Strogatz2000}, as:
\begin{equation}
	\dot{\theta}_i   = \epsilon_i - \frac{K}{N} \sum_{j=1}^{N} a_{ij} \sin(\theta_j - \theta_i),
\end{equation}
where $\epsilon_i $ is the frequency offset from the mean of the oscillator at node $i$, $\theta_i$ is the oscillator phase defined in terms of accumulated phase and an offset $\theta_i(t) = \epsilon_it + \phi_i$. $K$ is the coupling value, $a_{ij}$ are entries from the adjacency matrix of the graph. Notice that the coupling enters with a $-K/N$ prefactor, which makes the emergent state the ground state of the network and weights each edge by the factor $K$ renormalized by the number of oscillators in the network $N$. The nonlinearity comes from the $\sin(\theta_j - \theta_i)$-dependent coupling, which favors minimization of the phase differences. The $\sin(\theta_j - \theta_i)$ term introduces pairwise couplings that, in the case of inter-subgraph edges in the product graph, correspond to phase differences between one pair of QL bit subgraphs. For example, $a_1b_1c_1$ couples to the  subgraph $a_2b_1c_1$, introducing the phase difference between $a_1$ and $a_2$ (nominally the edge bias in the QL bit). See Fig. \ref{figExistQuad}a.

Introducing edges into the zero-coupling blocks suggests, concpetually, that higher-order (than pairwise) correlations can be added to the classical model to produce non-separable states. For example, by coupling $a_1b_1c_1$  to the  subgraph $a_2b_2c_1$, Fig. \ref{figExistQuad}b. These correlations come from additional coupling terms that can be added to the Kuramoto model\cite{Kuramotohigher1, Kuramotohigher2}:
\begin{equation}
	\dot{\theta}_i   = \epsilon_i - \frac{K}{N} \sum_{j=1}^{N} a_{ij} \sin(\theta_j - \theta_i) - \frac{K'}{N^3} \sum_{j=1}^{N} \sum_{l=1}^{N}  \sum_{m=1}^{N}  c_{ijlm} \sin(\theta_j - \theta_i + \theta_l - \theta_m),
\end{equation}
where $c_{ijlm}$ indicates an entry from the 3-simplex adjacency tensor. Simply adding edges to a network is obviously not the appropriate systematic way to include many-body correlations. The correlations come from variational coupling among a heirarchy of adjacency tensors. Furthermore, the resulting states may be challenging to analyze because classical many-body correlations are known to be difficult to distinguish from quantum correlations owing to the exponential complexity of a suitable measure for large state spaces. 

Numerical calculations provide superficial examples of what would happen if we `flatten' this high-dimensioanl structure and simply add more edges to the adjacency matrix, and weight the existing edges. For each calculation the product of two QL graphs is produced. For each (identical) QL graph there are 60 vertices and the graph is overall 40-regular random. All edges have bias of 1.  First note that the emergent state of the QL state thus produced is $v_1 = 0.5(|a_1b_1\rangle + |a_2b_2\rangle + |a_1b_2\rangle + |a_2b_1\rangle )$. Concurrence is a measure that, if non-zero, detects non-separable states\cite{Concurrence, Horodecki2009}. The concurrence calculated for this pure separable state $v_1$ is 0, as expected. 

Now add a coupling edge between the subgraphs $a_1b_2$ and $a_2b_1$. These vertices are never coupled by an edge in the Cartesian product of QL bit graphs. It is this new edge that renders the emergent state non-separable because the underlying graph cannot be composed using a Cartesian product. Then we find $v_2 = 0.43(|a_1b_1\rangle + |a_2b_2\rangle) + 0.56(|a_1b_2\rangle + |a_2b_1\rangle )$, with concurrence = 0.26. Taking the approach, but weighting the coupling edges inherited from the separable state by the scalar 0.01 and assigning coupling edges between the subgraphs $a_1b_2$ and $a_2b_1$ to have the value 1, we obtain $v_3 = 0.13(|a_1b_1\rangle + |a_2b_2\rangle) + 0.70(|a_1b_2\rangle + |a_2b_1\rangle )$, with concurrence = 0.93. Of course, this is for illustration only; the method for adding many-body phase correlations is not rigorous. We hypothesize that the QL model cannot be simply extended to force the states to be non-separable.

Notice that if we are to produce  \emph{maximally} entangled states, we would need to weight the  coupling edges inherited from the separable state by 0---that is, remove them, then add the new coupling edge(s). This is precisely what the action of a CNOT gate accomplishes\cite{Amati1} because, for the two QL bit case, it moves all the coupling edges into the zero-coupling blocks, see Appendix B. Without some reference point, once the graphs are disconnected it is unclear how the product basis is naturally defined for an arbitrary classical system. For example, although the subgraphs may be fortuitously synchronized, in general their relative phase is random. Whereas the product basis in the classical model is defined by the phase relationships between pairs of subgraphs. This observation shows why it is unlikely that entanglement can be generated by classical systems, although non-entangled, non-separable states are possible.

To summarize this section, it is known that classical non-separable states exist, as well-studied for classical light. Owing to the way superpositions in one QL bit need to depend, functionally, on superpositions in another QL bit in order to produce non-separable states, it is difficult to explore these states systematically in the QL graph model. The QL graph model is therefore better suited for producing separable states (it can generate all of them) and thereby suggest ways that coherence can be used as a resource.

\section{Outlook and some open questions}

We showed in Appendix A how the graph product can be optimized to produce a more compact graph with the essential properties required to generate states that mimic many of the properties of separable quantum states.  This optimized product representation gives a concrete visualization of the correlation structure in a QL state space and, interestingly, mirrors the structure of Boolean logic (because it is implicit in the poset structure\cite{Stanley1}). While entanglement cannot be produced by these classical structures, coherence can be leveraged as a resource. Moreover, the nonlinearity available to classical systems can produce very robust superposition states\cite{ScholesEntropy}. The open question is how such network structure offers new avenues for function compared to other, well-studied, networks\cite{Barabasi}?

Demonstrating how to exploit the QL states for computational or functional advantage is an open challenge\cite{Amati1}. Some points are worth noting. Regardless of QL phenomena, the underlying classical network incorporates intrinsic parallelism, like the universal memcomputing machines envisioned by Di Ventura and co-workers\cite{DiVentura1}. These machines have have the capacity to solve NP-complete problems in polynomial time\cite{DiVentura2}. Coupled oscillator networks are a compelling example of a system that can be realized as a QL resource, and such systems have already been developed for computing\cite{Csaba2020}. These platforms could certainly be adapted to the network architecture designed to produce QL states. All these known examples use the classical network for computation. The realization of a map to a QL state space opens up the possibility that the emergent states can also be used as a computational resource, which could allow classical circuits to perform calculations in a similar way to quantum computers\cite{Amati1}, but using the resource of superpositions.

We can ask, more speculatively, whether the  structure of basis states generated by the graph product be used to give advantages for function or computation in more abstract physical systems? Could such function be feasible in biological\cite{Whatis} or soft-matter systems\cite{slime1} by exploiting a manifestation of phase topology and scale-free connections? Or even for QL decision making\cite{Pothos2022, Khrennikov2023book, Ozawa2020, Busemeyer2014}?

The development of experiments to probe some of these questions will be important. For instance, experiments that compare measurements on the space of physical objects (the graphs) compared to the state space (spectrum of the graph). This could be achieved by studying suitably designed networks\cite{QLsync}. Similarly, physical examples of classical systems templated by QL graphs could allow systematic exploration of measurements on the state space, potentially enabling new structures of classical correlation to be exploited.

\section{Conclusions}

The main purpose of the present paper was to show that composite QL systems can closely mimic the separable states of quantum systems, and that suitable physical systems that exhibit these states exist. The QL graph representation of classical systems, developed in prior work\cite{ScholesQLstates, QLproducts}, was used to refine the search for classical systems that display states that have the same structure, of superpositions of states in a tensor product basis, as those of quantum systems, thus defining a general framework. A graph provides both a physical model for a classical system as well as defining the states, via the adjacency matrix of the graph.  It was shown that QL graphs can closely emulate separable states of composite quantum systems, such as coupled two-level systems. Moreover, we conclude that physical classical systems can be found that show these states. Examples include multipole moments of waves or networks of phase oscillators.


\appendix

\section{Edge topology of separable QL states}

The topology of a composite QL graph is revealed by reducing the graph complexity. Here it is shown how to optimize the Cartesian product of QL bit graphs in a way that keeps the eigenvalue of the emergent state constant as a function of products, rather than a multiple of $d$. Then the edge topology that connects subgraphs is highlighted.

The graph Cartesian product increases the vertex set multiplicatively---that is, the product of $q$ QL bits, each comprising $N$ vertices, generates a product graph comprising $N^q$ vertices. Thus, the size of the resource (the graph) scales proportionally with the size of the state space. Since each QL bit might contain many vertices, it is desirable to form the product most economically. A minimal product structure can be produced by contracting the graph as follows. This optimization of the product highlights the important design principle for a corresponding classical system, which is the topological of phase relationships. 

For simplicity, assume that each QL bit graph is identical. However, the precise structures do not matter in the end because the important correlation structure comes from the edges that connect the subgraphs, not the structure of the subgraphs themselves. As described above, forming the product $G_A \Box G_B$ involves installing a copy of $G_A$ at every vertex of $G_B$, or \emph{vice versa}, then connecting the vertices among $G_A$ graphs as prescribed by the edge structure in $G_B$. 

\begin{figure}
	\includegraphics[width=13.5 cm]{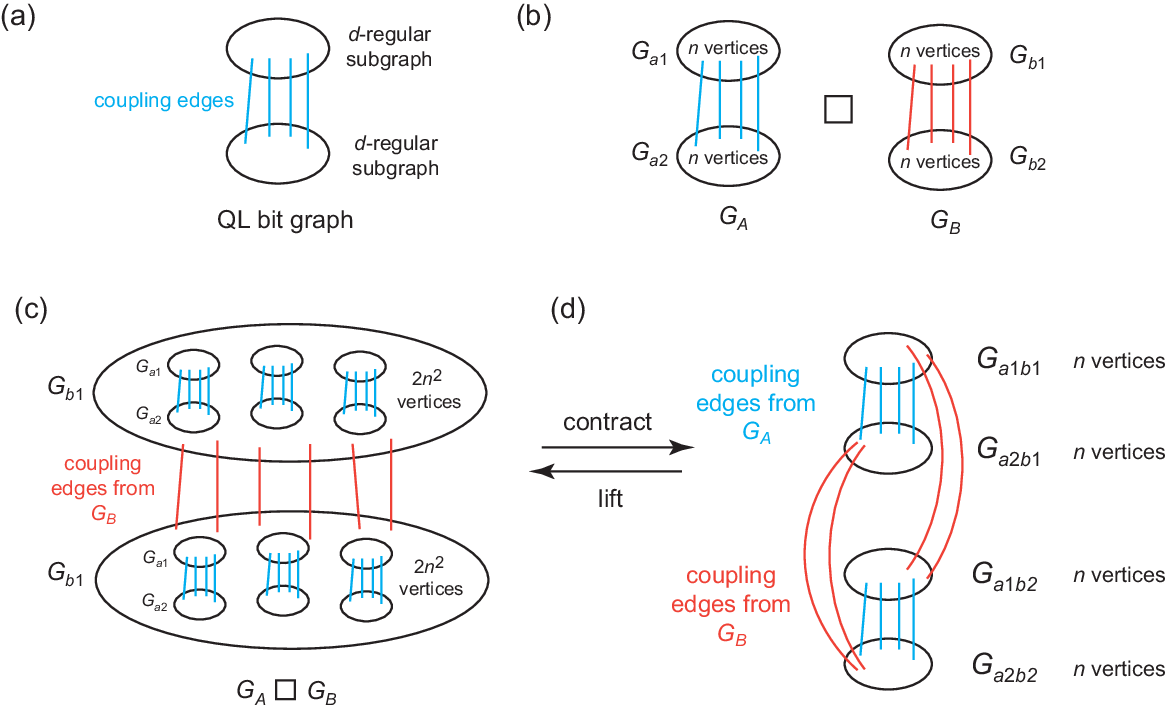}
	\caption{Outline of how the Cartesian product of QL bits, where each $d$-regular subgraph comprises $n$ vertices, produces a much larger graph, where each subgraph comprises $n^2$ vertices and is $2d$-regular. This large graph can be contracted to an optimal graph, where each $d$-regular subgraph comprises $n$ vertices, that retains the qualitative features of the large graph. }
	\label{figSync5}
\end{figure}

Let's view the product construction for QL bits schematically, Fig. \ref{figSync5}a,b. Recall that each subgraph $G_{a1}$, $G_{a2}$, $G_{b1}$, $G_{b2}$ is $d$-regular. Let's say each subgraph contains $n$ vertices. Notice that after we connect all the copies of $G_A$ that are associated with $G_{b1}$ (or similarly $G_{b2}$), we generate one large $2d$-regular subgraph on $n^2$ vertices by connecting all the $G_{a1}$ subgraphs together. We also produce one large $2d$-regular subgraph on $n^2$ vertices by connecting all the $G_{a2}$ subgraphs. This happens because we connect $n$ subgraphs and each contains $n$ vertices. 

Each vertex inherits the $d$ edges from the original graph (e.g. $G_{a1}$) and attains a further $d$ edges from the edge structure imposed by $G_{b1}$ (or $G_{b2}$), also $d$-regular. This increase in vertex degree is the reason that the eigenvalue of the emergent state of the uncontracted product $G_A \Box G_B$ is found at $2d$ instead of $d$\cite{QLproducts}.

Notice that within the vertex set labeled $G_{b1}$ (or $G_{b2}$) in the product, Fig. \ref{figSync5}c, that we now have a new $G_{a1}$ $2d$-regular subgraph connected by coupling edges to a new $G_{a2}$ $2d$-regular subgraph. We can simplify this graph without changing its properties qualitatively by contracting each $G_{a1}$ and each $G_{a2}$ $2d$-regular subgraph to corresponding $d$-regular sugbraphs. This produces an optimized product, Fig. \ref{figSync5}d. Conversely, the optimized product graph can be lifted to the full product using edge subdivisions. 

A subdivision of an edge connecting vertices $v_i$ and $v_j$ in a graph $G$ produces a new graph by replacing the edge with new connected vertices that connect to each of $v_i$ and $v_j$. These new connected vertices form a subgraph $X$, which might be a single vertex joining $v_i$ and $v_j$, or a more complex connected graph. A contraction of the subgraph $X$ in the graph reverses the subdivision, producing $G$. Similarly, we can apply a contraction to any subgraph in $G$. On the basis of these definitions and the properties of $d$-regular graphs (specifically the isoperimetric property), it is obvious how a $2d$-regular graph on $n^2$ vertices can be contracted to a $d$-regular graph on $n$ vertices. Divide the $n^2$ vertices into $n$ sets each containing $n$ vertices, then contract each of these sets into a single vertex. Concomitantly, the edges between sets collapse so that each contracted set, now a single vertex, has degree $d$. This  yields a $d$-regular graph on $n$ vertices. Conversely, subdivisions of edges of a $d$-regular graph can lift it to a $2d$-regular graph.

Now consider a state comprising $q$ QL-bits. Each QL bit contains $N = 2n$ vertices, $n$ in each subgraph. Therefore, $q$ QL bits contain $qN = 2qn$ vertices whereas their Cartesian product contains exponentially more vertices, $(2n)^q$ vertices in total. There are $2^q$ subgraphs, each comprising $n^q$ vertices. However, after contraction so that each subgraph contains $n$ vertices, we substantially lesson the high resource cost in subgraph vertices so that the overall graph now contains $n2^q$ vertices. 

The reason this dramatic reduction in resource requirement is possible, while still preserving the properties of the product, is that each subgraph is a large $d$-regular graph wherein the precise location of edges and number of vertices are unimportant. All that matters is that the graph is $d$-regular. The important edge construction derives from weakly connecting two $d$-regular subgraphs within each QL bit. From the definition of the Cartesian product of graphs, we have edges between subgraphs labeled $a_i$, $b_j$, $c_k, \dots$ to those labeled $a_l$, $b_m$, $c_n, \dots$ when either $i \neq l$ and $j = m$ and $k = n$, or when $i = l$ and $j \neq m$ and $k = n$ or when $i = l$ and $j = m$ and $k \neq n$. We preserve that important edge topology propagated through the product. Considering these principles, the QL bit graph products have a nice schematic representation, evident in Fig. \ref{figSync5}d.

Let's look at the structure of the optimized product graph in some more detail. The Cartesian graph product generates a new graph with a structure that gives the QL state space. There are two key ingredients underpinning this outcome. 

First, the subgraphs---that is, the $d$-regular graphs---are the vertices of the correlation structure shown in Fig. \ref{figSync5}d. These \emph{effective} vertices no longer correspond to the original subgraphs, that is, $G_{a1}$, $G_{a2}$, $G_{b1}$, and so on, that they were transcribed from. They now represent the product basis. We can read off those basis states for each subgraph (effective vertex), as shown in Fig. \ref{figSync5}d. Thus the effective vertices of the optimized product enumerate all permutations of the possible states of a set of QL bits---which is evident in the product basis.  We are free to choose whether the product function is symmetric or antisymmetric under pairwise permutations.

The second key ingredient is the edge structure---that is, the way the subgraphs are connected, shown as the colored edges in Fig. \ref{figSync5}d. Thhis edge structure, or topology of the graph, builds the state space by introducing nested correlations between graphs, analogous to the matrix representation of the analogous tensor product.  

We first provide some background and definitions needed for the sequel. A \emph{partially order set} (poset) $P$ comprises a collection of sets $s, t, u, \dots \in P$ together with a binary relation $\le$ that allows sets to be compared. $P$ satisfies axioms of reflexivity, antisymmetry, and transitivity (see Ref \cite{Stanley1} Chapter 3). The relevant example of a poset here is the Boolean poset $B_n$, defined as follows. Let $n \in \mathbb{N}$. The set of all $2^n$ subsets of $\{1, 2, \dots n \}$ becomes a poset by defining $s \le t$ in $B_n$ if the sets satisfy $s \subseteq t$. That is, the sets are ordered by inclusion.

For example, $\{ 1, 2\} \subseteq \{ 1, 2, 3\}$, which defines a kind of ordering such that these sets are said to be \emph{comparable}. On the other hand, $\{ 1, 2\}$ is \emph{incomparable} to $\{ 1, 3\}$ since neither is a subset of the other---yet both sets are included in $B_n, n \ge 3$. The collection of sets can be written as a Hasse diagram that enumerates all the sets and indicates those that are comparable by connecting lines. A \emph{chain} is a poset, or a subset of a poset, where any two elements are comparable. An \emph{antichain} is a subset of a poset where any two elements are incomparable. 

By considering together the two ingredients described above, we notice that the edges in the QL graphs introduce a partial ordering on the basis. That partial ordering is such that the optimized graph product representation generates the sets collected in a Hasse diagram for the $B_n$ posets. We define the ordering $a_1 \le a_2$, $b_1 \le b_2$, etc. That is, we equip the posets  for each QL bit with a rank function. A reason for the likeness of the Hasse diagram to any corresponding QL graph is that a new (graded) poset can be generated by a Cartesian product of posets, defined like our Cartesian product of graphs (see Sec. 3.2 of \cite{Stanley1}). The `edges' in the graph of posets (Hasse diagram) mean that the sets indicated at the vertices are comparable. Starting only with  the Hasse diagram for QL bits, we can form products and generate Hasse diagrams for the basis of the QL states, with a structure equivalent to that in the schematic graph representation.

\begin{figure}
	\includegraphics[width=13.5 cm]{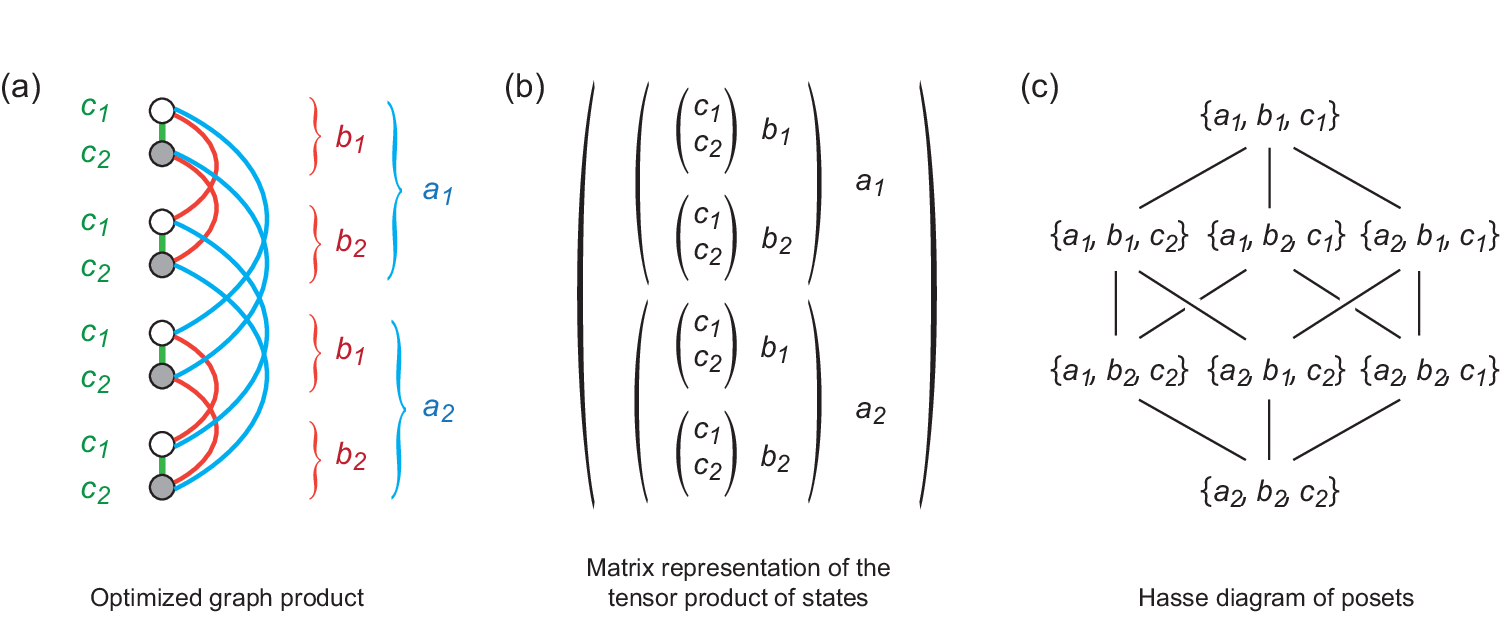}
	\caption{Comparison the graph product structure, tensor product construction of the basis of states, and the corresponding Hasse diagram for the set structure on the product states. }
	\label{figHasse}
\end{figure}   

In Fig. \ref{figHasse} we summarize this discussion by comparing the optimized product graph for the composite state formed from three QL bits, the corresponding matrix representation of the tensor product of the three QL bit emergent states, and the Hasse diagram that shows the partial order on the basis states imposed by the binary relation $\le$ defined above.The connections between partial sets in the Hasse diagram (the partial order) is equivalent to the edge structure in the optimized product graph.

\section{Maximally non-separable states require disconnected graphs}

For example, for the  product of two QL graphs recall that the graph for the composite system comprises four subgraphs, each linked by connecting edges (see Appendix). The subgraphs represent the basis states  $a_1b_1$,  $a_1b_2$,  $a_2b_1$, and  $a_2b_2$.  The corresponding adjacency matrix is
\begin{equation}
	A =
	\begin{pmatrix}
		0 & 1 & 1 & 0 \\
		1 & 0 & 0 & 1 \\
		1 & 0 & 0 & 1 \\
		0 & 1 & 1 & 0
	\end{pmatrix}.
\end{equation}
Using the definition for the CNOT gate unitary transformation on the adjacency matrix given in Ref. \cite{Amati1},
\begin{equation}
	U_{\text{CNOT}} =
	\begin{pmatrix}
		1 & 0 & 0& 0 \\
		0 & 0 & 0 & 1 \\
		0 & 0 & 1& 0 \\
		0 & 1 & 0 & 0
	\end{pmatrix}.
\end{equation}
The transformation $U_{\text{CNOT}} A U^{-1}_{\text{CNOT}}$ gives the new adjacency matrix
\begin{equation}
	A' =
	\begin{pmatrix}
		0 & 0 & 1 & 1 \\
		0 & 0 & 1 & 1 \\
		1 & 1 & 0 & 0 \\
		1 & 1 & 0 & 0
	\end{pmatrix}.
\end{equation}
where here the basis has been re-ordered to $a_1b_1$,  $a_2b_2$,  $a_2b_1$, and  $a_1b_2$. 

Notice that now the graph for the composite state becomes disconnected. After applying a CNOT gate, we find that subgraphs $a_1b_2$ and $a_2b_1$ are connected to each other, and similarly $a_1b_1$ and $a_2b_2$ become connected.  But these two units are not connected to each other because all the original edges have been deleted.

\begin{acknowledgments}
This research was funded by the National Science Foundation under Grant No. CHE-2537080.
\end{acknowledgments}


\vspace{6pt} 




\bibliography{Scholes_bib_Sept2025}

\begin{thebibliography}{81}%
\makeatletter
\providecommand \@ifxundefined [1]{%
 \@ifx{#1\undefined}
}%
\providecommand \@ifnum [1]{%
 \ifnum #1\expandafter \@firstoftwo
 \else \expandafter \@secondoftwo
 \fi
}%
\providecommand \@ifx [1]{%
 \ifx #1\expandafter \@firstoftwo
 \else \expandafter \@secondoftwo
 \fi
}%
\providecommand \natexlab [1]{#1}%
\providecommand \enquote  [1]{``#1''}%
\providecommand \bibnamefont  [1]{#1}%
\providecommand \bibfnamefont [1]{#1}%
\providecommand \citenamefont [1]{#1}%
\providecommand \href@noop [0]{\@secondoftwo}%
\providecommand \href [0]{\begingroup \@sanitize@url \@href}%
\providecommand \@href[1]{\@@startlink{#1}\@@href}%
\providecommand \@@href[1]{\endgroup#1\@@endlink}%
\providecommand \@sanitize@url [0]{\catcode `\\12\catcode `\$12\catcode
  `\&12\catcode `\#12\catcode `\^12\catcode `\_12\catcode `\%12\relax}%
\providecommand \@@startlink[1]{}%
\providecommand \@@endlink[0]{}%
\providecommand \url  [0]{\begingroup\@sanitize@url \@url }%
\providecommand \@url [1]{\endgroup\@href {#1}{\urlprefix }}%
\providecommand \urlprefix  [0]{URL }%
\providecommand \Eprint [0]{\href }%
\providecommand \doibase [0]{https://doi.org/}%
\providecommand \selectlanguage [0]{\@gobble}%
\providecommand \bibinfo  [0]{\@secondoftwo}%
\providecommand \bibfield  [0]{\@secondoftwo}%
\providecommand \translation [1]{[#1]}%
\providecommand \BibitemOpen [0]{}%
\providecommand \bibitemStop [0]{}%
\providecommand \bibitemNoStop [0]{.\EOS\space}%
\providecommand \EOS [0]{\spacefactor3000\relax}%
\providecommand \BibitemShut  [1]{\csname bibitem#1\endcsname}%
\let\auto@bib@innerbib\@empty
\bibitem [{\citenamefont {Horodecki}\ \emph {et~al.}(2009)\citenamefont
  {Horodecki}, \citenamefont {Horodecki}, \citenamefont {Horodecki},\ and\
  \citenamefont {Horodecki}}]{Horodecki2009}%
  \BibitemOpen
  \bibfield  {author} {\bibinfo {author} {\bibfnamefont {R.}~\bibnamefont
  {Horodecki}}, \bibinfo {author} {\bibfnamefont {P.}~\bibnamefont
  {Horodecki}}, \bibinfo {author} {\bibfnamefont {M.}~\bibnamefont
  {Horodecki}},\ and\ \bibinfo {author} {\bibfnamefont {K.}~\bibnamefont
  {Horodecki}},\ }\bibfield  {title} {\bibinfo {title} {Quantum entanglement},\
  }\href@noop {} {\bibfield  {journal} {\bibinfo  {journal} {Rev. Mod. Phys.}\
  }\textbf {\bibinfo {volume} {81}},\ \bibinfo {pages} {865} (\bibinfo {year}
  {2009})}\BibitemShut {NoStop}%
\bibitem [{\citenamefont {Nielsen}\ and\ \citenamefont
  {Chuang}(2016)}]{NielsenChuang}%
  \BibitemOpen
  \bibfield  {author} {\bibinfo {author} {\bibfnamefont {M.~A.}\ \bibnamefont
  {Nielsen}}\ and\ \bibinfo {author} {\bibfnamefont {I.~L.}\ \bibnamefont
  {Chuang}},\ }\href@noop {} {\emph {\bibinfo {title} {Quantum Computation and
  Quantum Information}}}\ (\bibinfo  {publisher} {Cambridge University Press,
  Cambridge},\ \bibinfo {year} {2016})\BibitemShut {NoStop}%
\bibitem [{\citenamefont {Huelga}\ and\ \citenamefont
  {Plenio}(2013)}]{PlenioQBiol}%
  \BibitemOpen
  \bibfield  {author} {\bibinfo {author} {\bibfnamefont {S.~F.}\ \bibnamefont
  {Huelga}}\ and\ \bibinfo {author} {\bibfnamefont {M.~B.}\ \bibnamefont
  {Plenio}},\ }\bibfield  {title} {\bibinfo {title} {Vibrations, quanta and
  biology},\ }\href@noop {} {\bibfield  {journal} {\bibinfo  {journal}
  {Contemporary Physics}\ }\textbf {\bibinfo {volume} {54}},\ \bibinfo {pages}
  {181} (\bibinfo {year} {2013})}\BibitemShut {NoStop}%
\bibitem [{\citenamefont {Scholes}\ and\ \citenamefont
  {Fleming}(2026)}]{Whatis}%
  \BibitemOpen
  \bibfield  {author} {\bibinfo {author} {\bibfnamefont {G.~D.}\ \bibnamefont
  {Scholes}}\ and\ \bibinfo {author} {\bibfnamefont {G.~R.}\ \bibnamefont
  {Fleming}},\ }\bibfield  {title} {\bibinfo {title} {What is quantum
  biology?},\ }\href@noop {} {\bibfield  {journal} {\bibinfo  {journal} {Proc.
  Natl. Acad. Sci. (USA)}\ }\textbf {\bibinfo {volume} {123}},\ \bibinfo
  {pages} {e2531134123} (\bibinfo {year} {2026})}\BibitemShut {NoStop}%
\bibitem [{\citenamefont {Scholes}(2024)}]{ScholesQLstates}%
  \BibitemOpen
  \bibfield  {author} {\bibinfo {author} {\bibfnamefont {G.~D.}\ \bibnamefont
  {Scholes}},\ }\bibfield  {title} {\bibinfo {title} {Quantum-like states on
  complex synchronized networks},\ }\href@noop {} {\bibfield  {journal}
  {\bibinfo  {journal} {Proc. R. Soc. A}\ }\textbf {\bibinfo {volume} {480}},\
  \bibinfo {pages} {20240209} (\bibinfo {year} {2024})}\BibitemShut {NoStop}%
\bibitem [{\citenamefont {Scholes}\ and\ \citenamefont
  {Amati}(2025)}]{QLproducts}%
  \BibitemOpen
  \bibfield  {author} {\bibinfo {author} {\bibfnamefont {G.~D.}\ \bibnamefont
  {Scholes}}\ and\ \bibinfo {author} {\bibfnamefont {G.}~\bibnamefont
  {Amati}},\ }\bibfield  {title} {\bibinfo {title} {Quantum-like product states
  constructed from classical networks},\ }\href@noop {} {\bibfield  {journal}
  {\bibinfo  {journal} {Phys. Rev. Lett.}\ }\textbf {\bibinfo {volume} {134}},\
  \bibinfo {pages} {060202} (\bibinfo {year} {2025})}\BibitemShut {NoStop}%
\bibitem [{\citenamefont {Scholes}(2026)}]{QLsync}%
  \BibitemOpen
  \bibfield  {author} {\bibinfo {author} {\bibfnamefont {G.~D.}\ \bibnamefont
  {Scholes}},\ }\bibfield  {title} {\bibinfo {title} {Dynamics in an emergent
  quantum-like state space generated by a nonlinear classical network},\
  }\href@noop {} {\bibfield  {journal} {\bibinfo  {journal} {Physica Scripta}\
  }\textbf {\bibinfo {volume} {101}},\ \bibinfo {pages} {055205} (\bibinfo
  {year} {2026})}\BibitemShut {NoStop}%
\bibitem [{\citenamefont {Saha}\ and\ \citenamefont
  {Scholes}(2025)}]{Debadrita1}%
  \BibitemOpen
  \bibfield  {author} {\bibinfo {author} {\bibfnamefont {D.}~\bibnamefont
  {Saha}}\ and\ \bibinfo {author} {\bibfnamefont {G.}~\bibnamefont {Scholes}},\
  }\bibfield  {title} {\bibinfo {title} {Multi-dimensional quantum like
  resources from complex synchronized networks},\ }\href@noop {} {\bibfield
  {journal} {\bibinfo  {journal} {Entropy}\ }\textbf {\bibinfo {volume} {17}},\
  \bibinfo {pages} {963} (\bibinfo {year} {2025})}\BibitemShut {NoStop}%
\bibitem [{\citenamefont {Amati}\ and\ \citenamefont
  {Scholes}(2025{\natexlab{a}})}]{Amati1}%
  \BibitemOpen
  \bibfield  {author} {\bibinfo {author} {\bibfnamefont {G.}~\bibnamefont
  {Amati}}\ and\ \bibinfo {author} {\bibfnamefont {G.~D.}\ \bibnamefont
  {Scholes}},\ }\bibfield  {title} {\bibinfo {title} {Quantum information with
  quantum-like bits},\ }\href@noop {} {\bibfield  {journal} {\bibinfo
  {journal} {Phys. Rev. A}\ }\textbf {\bibinfo {volume} {111}},\ \bibinfo
  {pages} {062203} (\bibinfo {year} {2025}{\natexlab{a}})}\BibitemShut
  {NoStop}%
\bibitem [{\citenamefont {Amati}\ and\ \citenamefont
  {Scholes}(2025{\natexlab{b}})}]{Amati2}%
  \BibitemOpen
  \bibfield  {author} {\bibinfo {author} {\bibfnamefont {G.}~\bibnamefont
  {Amati}}\ and\ \bibinfo {author} {\bibfnamefont {G.~D.}\ \bibnamefont
  {Scholes}},\ }\bibfield  {title} {\bibinfo {title} {Encoding quantum-like
  information in classical synchronizing dynamics},\ }\href@noop {} {\bibfield
  {journal} {\bibinfo  {journal} {Phys. Rev. A}\ }\textbf {\bibinfo {volume}
  {112}},\ \bibinfo {pages} {032423} (\bibinfo {year}
  {2025}{\natexlab{b}})}\BibitemShut {NoStop}%
\bibitem [{\citenamefont {Khrennikov}(2003)}]{Khrennikov2003}%
  \BibitemOpen
  \bibfield  {author} {\bibinfo {author} {\bibfnamefont {A.}~\bibnamefont
  {Khrennikov}},\ }\bibfield  {title} {\bibinfo {title} {Quantum-like formalism
  for cognitive measurements},\ }\href@noop {} {\bibfield  {journal} {\bibinfo
  {journal} {BioSystems}\ }\textbf {\bibinfo {volume} {70}},\ \bibinfo {pages}
  {211} (\bibinfo {year} {2003})}\BibitemShut {NoStop}%
\bibitem [{\citenamefont {Khrennikov}(2005{\natexlab{a}})}]{Khrennikov2005}%
  \BibitemOpen
  \bibfield  {author} {\bibinfo {author} {\bibfnamefont {A.~Y.}\ \bibnamefont
  {Khrennikov}},\ }\bibfield  {title} {\bibinfo {title} {Interference in the
  classical probabilistic model and its representation in complex hilbert
  space},\ }\href@noop {} {\bibfield  {journal} {\bibinfo  {journal} {Physica
  E}\ }\textbf {\bibinfo {volume} {29}},\ \bibinfo {pages} {226} (\bibinfo
  {year} {2005}{\natexlab{a}})}\BibitemShut {NoStop}%
\bibitem [{\citenamefont {Khrennikov}(2005{\natexlab{b}})}]{Khrennikov2005b}%
  \BibitemOpen
  \bibfield  {author} {\bibinfo {author} {\bibfnamefont {A.}~\bibnamefont
  {Khrennikov}},\ }\bibfield  {title} {\bibinfo {title} {Interference of
  probabilities in the classical probabilistic framework},\ }\href@noop {}
  {\bibfield  {journal} {\bibinfo  {journal} {Fuzzy Sets Sys.}\ }\textbf
  {\bibinfo {volume} {155}},\ \bibinfo {pages} {4} (\bibinfo {year}
  {2005}{\natexlab{b}})}\BibitemShut {NoStop}%
\bibitem [{\citenamefont {Khrennikov}(2010)}]{Khrennikov1}%
  \BibitemOpen
  \bibfield  {author} {\bibinfo {author} {\bibfnamefont {A.}~\bibnamefont
  {Khrennikov}},\ }\href@noop {} {\emph {\bibinfo {title} {Ubiquitous Quantum
  Structure}}}\ (\bibinfo  {publisher} {Springer, Heidelberg},\ \bibinfo {year}
  {2010})\BibitemShut {NoStop}%
\bibitem [{\citenamefont {Ozawa}\ and\ \citenamefont
  {Khrennikov}(2020)}]{Ozawa2020}%
  \BibitemOpen
  \bibfield  {author} {\bibinfo {author} {\bibfnamefont {M.}~\bibnamefont
  {Ozawa}}\ and\ \bibinfo {author} {\bibfnamefont {A.}~\bibnamefont
  {Khrennikov}},\ }\bibfield  {title} {\bibinfo {title} {Application of theory
  of quantum instruments to psychology: Combination of question order effect
  with response replicability effect},\ }\href@noop {} {\bibfield  {journal}
  {\bibinfo  {journal} {Entropy}\ }\textbf {\bibinfo {volume} {22}},\ \bibinfo
  {pages} {37} (\bibinfo {year} {2020})}\BibitemShut {NoStop}%
\bibitem [{\citenamefont {Khrennikov}\ and\ \citenamefont
  {Basieva}(2014)}]{KB2013}%
  \BibitemOpen
  \bibfield  {author} {\bibinfo {author} {\bibfnamefont {A.}~\bibnamefont
  {Khrennikov}}\ and\ \bibinfo {author} {\bibfnamefont {I.}~\bibnamefont
  {Basieva}},\ }\bibfield  {title} {\bibinfo {title} {Quantum model for
  psychological measurements: From the projection postulate to interference of
  mental observables represented as positive operator valued measures},\
  }\href@noop {} {\bibfield  {journal} {\bibinfo  {journal} {NeuroQuantology}\
  }\textbf {\bibinfo {volume} {12}},\ \bibinfo {pages} {324} (\bibinfo {year}
  {2014})}\BibitemShut {NoStop}%
\bibitem [{\citenamefont {Khrennikov}(2016)}]{Khrennikov2016}%
  \BibitemOpen
  \bibfield  {author} {\bibinfo {author} {\bibfnamefont {A.}~\bibnamefont
  {Khrennikov}},\ }\bibfield  {title} {\bibinfo {title} {Quantum bayesianism as
  the basis of general theory of decision-making},\ }\href@noop {} {\bibfield
  {journal} {\bibinfo  {journal} {Phil. Trans. R. Soc. A}\ }\textbf {\bibinfo
  {volume} {374}},\ \bibinfo {pages} {20150245} (\bibinfo {year}
  {2016})}\BibitemShut {NoStop}%
\bibitem [{\citenamefont {Khrennikov}\ \emph {et~al.}(2018)\citenamefont
  {Khrennikov}, \citenamefont {Basieva}, \citenamefont {Pothos},\ and\
  \citenamefont {Yamato}}]{Khrennikov2018SciRep}%
  \BibitemOpen
  \bibfield  {author} {\bibinfo {author} {\bibfnamefont {A.}~\bibnamefont
  {Khrennikov}}, \bibinfo {author} {\bibfnamefont {I.}~\bibnamefont {Basieva}},
  \bibinfo {author} {\bibfnamefont {E.~M.}\ \bibnamefont {Pothos}},\ and\
  \bibinfo {author} {\bibfnamefont {I.}~\bibnamefont {Yamato}},\ }\bibfield
  {title} {\bibinfo {title} {Quantum probability in decision making from
  quantum information representation of neuronal states},\ }\href@noop {}
  {\bibfield  {journal} {\bibinfo  {journal} {Sci. Rep.}\ }\textbf {\bibinfo
  {volume} {8}},\ \bibinfo {pages} {16225} (\bibinfo {year}
  {2018})}\BibitemShut {NoStop}%
\bibitem [{\citenamefont {Khrennikov}(2023)}]{Khrennikov2023book}%
  \BibitemOpen
  \bibfield  {author} {\bibinfo {author} {\bibfnamefont {A.~Y.}\ \bibnamefont
  {Khrennikov}},\ }\href@noop {} {\emph {\bibinfo {title} {Open Quantum Systems
  in Biology, Cognitive and Social Sciences}}}\ (\bibinfo  {publisher}
  {Springer Nature, Berlin},\ \bibinfo {year} {2023})\BibitemShut {NoStop}%
\bibitem [{\citenamefont {Elze}(2017)}]{Elze2017}%
  \BibitemOpen
  \bibfield  {author} {\bibinfo {author} {\bibfnamefont {H.-T.}\ \bibnamefont
  {Elze}},\ }\bibfield  {title} {\bibinfo {title} {Quantum models as classical
  cellular automata},\ }\href@noop {} {\bibfield  {journal} {\bibinfo
  {journal} {J. Phys.: Conf. Ser.}\ }\textbf {\bibinfo {volume} {845}},\
  \bibinfo {pages} {012022} (\bibinfo {year} {2017})}\BibitemShut {NoStop}%
\bibitem [{\citenamefont {Elze}(2024)}]{Elze2024}%
  \BibitemOpen
  \bibfield  {author} {\bibinfo {author} {\bibfnamefont {H.-T.}\ \bibnamefont
  {Elze}},\ }\bibfield  {title} {\bibinfo {title} {Cellular automaton ontology,
  bits, qubits and the dirac equation},\ }\href@noop {} {\bibfield  {journal}
  {\bibinfo  {journal} {Int. J. Quantum Info.}\ }\textbf {\bibinfo {volume}
  {22}},\ \bibinfo {pages} {2450013} (\bibinfo {year} {2024})}\BibitemShut
  {NoStop}%
\bibitem [{\citenamefont {Couder}\ and\ \citenamefont {Fort}(2006)}]{Fort2006}%
  \BibitemOpen
  \bibfield  {author} {\bibinfo {author} {\bibfnamefont {Y.}~\bibnamefont
  {Couder}}\ and\ \bibinfo {author} {\bibfnamefont {E.}~\bibnamefont {Fort}},\
  }\bibfield  {title} {\bibinfo {title} {Single-particle diffraction and
  interference at a macroscopic scale},\ }\href@noop {} {\bibfield  {journal}
  {\bibinfo  {journal} {Phys. Rev. Lett.}\ }\textbf {\bibinfo {volume} {97}},\
  \bibinfo {pages} {154101} (\bibinfo {year} {2006})}\BibitemShut {NoStop}%
\bibitem [{\citenamefont {Scholes}(2025)}]{Excitonics2024}%
  \BibitemOpen
  \bibfield  {author} {\bibinfo {author} {\bibfnamefont {G.~D.}\ \bibnamefont
  {Scholes}},\ }\bibfield  {title} {\bibinfo {title} {Graphs that predict
  exciton delocalization},\ }\href@noop {} {\bibfield  {journal} {\bibinfo
  {journal} {Phil. Trans.R. Soc. A}\ }\textbf {\bibinfo {volume} {accepted}},\
  \bibinfo {pages} {DOI:10.1098/rsta.2023.0250, arXiv:2501.09843} (\bibinfo
  {year} {2025})}\BibitemShut {NoStop}%
\bibitem [{\citenamefont {Zaslavsky}(1989)}]{Zaslavsky1}%
  \BibitemOpen
  \bibfield  {author} {\bibinfo {author} {\bibfnamefont {T.}~\bibnamefont
  {Zaslavsky}},\ }\bibfield  {title} {\bibinfo {title} {Biased graphs. i. bias,
  balance, and gains},\ }\href@noop {} {\bibfield  {journal} {\bibinfo
  {journal} {J. Comb. Theory Ser. B}\ }\textbf {\bibinfo {volume} {47}},\
  \bibinfo {pages} {32} (\bibinfo {year} {1989})}\BibitemShut {NoStop}%
\bibitem [{\citenamefont {Mehatari}\ \emph {et~al.}(2022)\citenamefont
  {Mehatari}, \citenamefont {Kannan},\ and\ \citenamefont
  {Samanta}}]{Mehatari}%
  \BibitemOpen
  \bibfield  {author} {\bibinfo {author} {\bibfnamefont {R.}~\bibnamefont
  {Mehatari}}, \bibinfo {author} {\bibfnamefont {M.~R.}\ \bibnamefont
  {Kannan}},\ and\ \bibinfo {author} {\bibfnamefont {A.}~\bibnamefont
  {Samanta}},\ }\bibfield  {title} {\bibinfo {title} {On the adjacency matrix
  of a complex unit gain graph},\ }\href@noop {} {\bibfield  {journal}
  {\bibinfo  {journal} {Lin. Multilin. Algebra}\ }\textbf {\bibinfo {volume}
  {70}},\ \bibinfo {pages} {1798} (\bibinfo {year} {2022})}\BibitemShut
  {NoStop}%
\bibitem [{\citenamefont {Reff}(2012)}]{Reff2012}%
  \BibitemOpen
  \bibfield  {author} {\bibinfo {author} {\bibfnamefont {N.}~\bibnamefont
  {Reff}},\ }\bibfield  {title} {\bibinfo {title} {Spectral properties of
  complex unit gain graphs},\ }\href@noop {} {\bibfield  {journal} {\bibinfo
  {journal} {Lin. Alg. Appl.}\ }\textbf {\bibinfo {volume} {436}},\ \bibinfo
  {pages} {3165} (\bibinfo {year} {2012})}\BibitemShut {NoStop}%
\bibitem [{\citenamefont {Chegnizadeh}\ \emph {et~al.}(2024)\citenamefont
  {Chegnizadeh}, \citenamefont {Scigliuzzo}, \citenamefont {Youssefi},
  \citenamefont {Kono}, \citenamefont {Guzovskii},\ and\ \citenamefont
  {Kippenberg}}]{Kippenberg2024}%
  \BibitemOpen
  \bibfield  {author} {\bibinfo {author} {\bibfnamefont {M.}~\bibnamefont
  {Chegnizadeh}}, \bibinfo {author} {\bibfnamefont {M.}~\bibnamefont
  {Scigliuzzo}}, \bibinfo {author} {\bibfnamefont {A.}~\bibnamefont
  {Youssefi}}, \bibinfo {author} {\bibfnamefont {S.}~\bibnamefont {Kono}},
  \bibinfo {author} {\bibfnamefont {E.}~\bibnamefont {Guzovskii}},\ and\
  \bibinfo {author} {\bibfnamefont {T.~J.}\ \bibnamefont {Kippenberg}},\
  }\bibfield  {title} {\bibinfo {title} {Quantum collective motion of
  macroscopic mechanical oscillators},\ }\href@noop {} {\bibfield  {journal}
  {\bibinfo  {journal} {Science}\ }\textbf {\bibinfo {volume} {386}},\ \bibinfo
  {pages} {1383} (\bibinfo {year} {2024})}\BibitemShut {NoStop}%
\bibitem [{\citenamefont {Sarnak}(2004)}]{Sarnak2004}%
  \BibitemOpen
  \bibfield  {author} {\bibinfo {author} {\bibfnamefont {P.}~\bibnamefont
  {Sarnak}},\ }\bibfield  {title} {\bibinfo {title} {What is an expander?},\
  }\href@noop {} {\bibfield  {journal} {\bibinfo  {journal} {Notices AMS}\
  }\textbf {\bibinfo {volume} {51}},\ \bibinfo {pages} {762} (\bibinfo {year}
  {2004})}\BibitemShut {NoStop}%
\bibitem [{\citenamefont {Krebbs}\ and\ \citenamefont
  {Shaheen}(2011)}]{expandersguide}%
  \BibitemOpen
  \bibfield  {author} {\bibinfo {author} {\bibfnamefont {M.}~\bibnamefont
  {Krebbs}}\ and\ \bibinfo {author} {\bibfnamefont {A.}~\bibnamefont
  {Shaheen}},\ }\href@noop {} {\emph {\bibinfo {title} {Expander Families and
  Cayley Graphs: A beginner’s guide}}}\ (\bibinfo  {publisher} {Oxford
  University Press, Oxford},\ \bibinfo {year} {2011})\BibitemShut {NoStop}%
\bibitem [{\citenamefont {Lubotsky}(2012)}]{Lubotzky}%
  \BibitemOpen
  \bibfield  {author} {\bibinfo {author} {\bibfnamefont {A.}~\bibnamefont
  {Lubotsky}},\ }\bibfield  {title} {\bibinfo {title} {Expander graphs in pure
  and applied mathematics},\ }\href@noop {} {\bibfield  {journal} {\bibinfo
  {journal} {Bull. Amer. Math. Soc.}\ }\textbf {\bibinfo {volume} {49}},\
  \bibinfo {pages} {113} (\bibinfo {year} {2012})}\BibitemShut {NoStop}%
\bibitem [{\citenamefont {Hoory}\ \emph {et~al.}(2006)\citenamefont {Hoory},
  \citenamefont {Linial},\ and\ \citenamefont {Wigderson}}]{Expanders}%
  \BibitemOpen
  \bibfield  {author} {\bibinfo {author} {\bibfnamefont {S.}~\bibnamefont
  {Hoory}}, \bibinfo {author} {\bibfnamefont {N.}~\bibnamefont {Linial}},\ and\
  \bibinfo {author} {\bibfnamefont {A.}~\bibnamefont {Wigderson}},\ }\bibfield
  {title} {\bibinfo {title} {Expander graphs and their applications},\
  }\href@noop {} {\bibfield  {journal} {\bibinfo  {journal} {Bull. Amer. Math.
  Soc.}\ }\textbf {\bibinfo {volume} {43}},\ \bibinfo {pages} {439} (\bibinfo
  {year} {2006})}\BibitemShut {NoStop}%
\bibitem [{\citenamefont {Alon}(2021)}]{Expanders2}%
  \BibitemOpen
  \bibfield  {author} {\bibinfo {author} {\bibfnamefont {N.}~\bibnamefont
  {Alon}},\ }\bibfield  {title} {\bibinfo {title} {Explicity expanders of every
  degree and size},\ }\href@noop {} {\bibfield  {journal} {\bibinfo  {journal}
  {Combinatorica}\ }\textbf {\bibinfo {volume} {41}},\ \bibinfo {pages} {447}
  (\bibinfo {year} {2021})}\BibitemShut {NoStop}%
\bibitem [{\citenamefont {Alon}(1986)}]{Alon1986}%
  \BibitemOpen
  \bibfield  {author} {\bibinfo {author} {\bibfnamefont {N.}~\bibnamefont
  {Alon}},\ }\bibfield  {title} {\bibinfo {title} {Eigenvalues and expanders},\
  }\href@noop {} {\bibfield  {journal} {\bibinfo  {journal} {Combinatorica}\
  }\textbf {\bibinfo {volume} {6}},\ \bibinfo {pages} {83} (\bibinfo {year}
  {1986})}\BibitemShut {NoStop}%
\bibitem [{\citenamefont {Tao}(2015)}]{Tao-expanders}%
  \BibitemOpen
  \bibfield  {author} {\bibinfo {author} {\bibfnamefont {T.}~\bibnamefont
  {Tao}},\ }\href@noop {} {\emph {\bibinfo {title} {Expansion in finite simple
  groups of Lie type}}}\ (\bibinfo  {publisher} {American Mathematical Society,
  Providence RI},\ \bibinfo {year} {2015})\BibitemShut {NoStop}%
\bibitem [{\citenamefont {Lee}\ \emph {et~al.}(2024)\citenamefont {Lee},
  \citenamefont {Kweun}, \citenamefont {Lee}, \citenamefont {Seung},\ and\
  \citenamefont {Kim}}]{Elastic}%
  \BibitemOpen
  \bibfield  {author} {\bibinfo {author} {\bibfnamefont {J.}~\bibnamefont
  {Lee}}, \bibinfo {author} {\bibfnamefont {M.~J.}\ \bibnamefont {Kweun}},
  \bibinfo {author} {\bibfnamefont {W.}~\bibnamefont {Lee}}, \bibinfo {author}
  {\bibfnamefont {H.~M.}\ \bibnamefont {Seung}},\ and\ \bibinfo {author}
  {\bibfnamefont {Y.~Y.}\ \bibnamefont {Kim}},\ }\bibfield  {title} {\bibinfo
  {title} {Perfect circular polarization of elastic waves in solid media},\
  }\href@noop {} {\bibfield  {journal} {\bibinfo  {journal} {Nature Commun.}\
  }\textbf {\bibinfo {volume} {15}},\ \bibinfo {pages} {992} (\bibinfo {year}
  {2024})}\BibitemShut {NoStop}%
\bibitem [{\citenamefont {Kliger}\ \emph {et~al.}(1990)\citenamefont {Kliger},
  \citenamefont {Lewis},\ and\ \citenamefont {Randall}}]{Kliger}%
  \BibitemOpen
  \bibfield  {author} {\bibinfo {author} {\bibfnamefont {D.}~\bibnamefont
  {Kliger}}, \bibinfo {author} {\bibfnamefont {J.}~\bibnamefont {Lewis}},\ and\
  \bibinfo {author} {\bibfnamefont {C.}~\bibnamefont {Randall}},\ }\href@noop
  {} {\emph {\bibinfo {title} {Polarized light in optics and spectroscopy}}}\
  (\bibinfo  {publisher} {Academic Press, New York},\ \bibinfo {year}
  {1990})\BibitemShut {NoStop}%
\bibitem [{\citenamefont {Stillwell}(2010)}]{NaiveLie}%
  \BibitemOpen
  \bibfield  {author} {\bibinfo {author} {\bibfnamefont {J.}~\bibnamefont
  {Stillwell}},\ }\href@noop {} {\emph {\bibinfo {title} {Naive Lie Theory}}}\
  (\bibinfo  {publisher} {Springer, Undergraduate Texts in Mathematics, New
  York},\ \bibinfo {year} {2010})\BibitemShut {NoStop}%
\bibitem [{\citenamefont {Cohn}(1968)}]{Cohn}%
  \BibitemOpen
  \bibfield  {author} {\bibinfo {author} {\bibfnamefont {P.}~\bibnamefont
  {Cohn}},\ }\href@noop {} {\emph {\bibinfo {title} {Lie Groups}}}\ (\bibinfo
  {publisher} {Cambridge University Press, Cambridge},\ \bibinfo {year}
  {1968})\BibitemShut {NoStop}%
\bibitem [{\citenamefont {Hall}(2015)}]{Hall2015}%
  \BibitemOpen
  \bibfield  {author} {\bibinfo {author} {\bibfnamefont {B.~C.}\ \bibnamefont
  {Hall}},\ }\href@noop {} {\emph {\bibinfo {title} {Lie Groups, Lie Algebras,
  and Representations}}}\ (\bibinfo  {publisher} {Springer},\ \bibinfo {year}
  {2015})\BibitemShut {NoStop}%
\bibitem [{\citenamefont {Woit}(2017)}]{Woit}%
  \BibitemOpen
  \bibfield  {author} {\bibinfo {author} {\bibfnamefont {P.}~\bibnamefont
  {Woit}},\ }\href@noop {} {\emph {\bibinfo {title} {Quantum theory, groups and
  representations. An introduction}}}\ (\bibinfo  {publisher} {Springer, Cham,
  Switzerland},\ \bibinfo {year} {2017})\BibitemShut {NoStop}%
\bibitem [{\citenamefont {Imrich}\ and\ \citenamefont
  {Klav\u{z}ar}(2000)}]{GraphProducts}%
  \BibitemOpen
  \bibfield  {author} {\bibinfo {author} {\bibfnamefont {W.}~\bibnamefont
  {Imrich}}\ and\ \bibinfo {author} {\bibfnamefont {S.}~\bibnamefont
  {Klav\u{z}ar}},\ }\href@noop {} {\emph {\bibinfo {title} {Product Graphs:
  Structure and Recognition}}}\ (\bibinfo  {publisher} {Wiley, New York},\
  \bibinfo {year} {2000})\BibitemShut {NoStop}%
\bibitem [{\citenamefont {Sabidussi}(1960)}]{Sabidussi}%
  \BibitemOpen
  \bibfield  {author} {\bibinfo {author} {\bibfnamefont {G.}~\bibnamefont
  {Sabidussi}},\ }\bibfield  {title} {\bibinfo {title} {Graph multiplication},\
  }\href@noop {} {\bibfield  {journal} {\bibinfo  {journal} {Math. Zeitschr.}\
  }\textbf {\bibinfo {volume} {72}},\ \bibinfo {pages} {446} (\bibinfo {year}
  {1960})}\BibitemShut {NoStop}%
\bibitem [{\citenamefont {Scholes}(2020)}]{Scholes2020}%
  \BibitemOpen
  \bibfield  {author} {\bibinfo {author} {\bibfnamefont {G.~D.}\ \bibnamefont
  {Scholes}},\ }\bibfield  {title} {\bibinfo {title} {Polaritons and excitons:
  Hamiltonian design for enhanced coherence},\ }\href@noop {} {\bibfield
  {journal} {\bibinfo  {journal} {Proc. R. Soc. A}\ }\textbf {\bibinfo {volume}
  {476}},\ \bibinfo {pages} {20200278} (\bibinfo {year} {2020})}\BibitemShut
  {NoStop}%
\bibitem [{\citenamefont {Dong}\ \emph {et~al.}(2021)\citenamefont {Dong},
  \citenamefont {Juricic},\ and\ \citenamefont {Roy}}]{topcircuit2}%
  \BibitemOpen
  \bibfield  {author} {\bibinfo {author} {\bibfnamefont {J.}~\bibnamefont
  {Dong}}, \bibinfo {author} {\bibfnamefont {V.}~\bibnamefont {Juricic}},\ and\
  \bibinfo {author} {\bibfnamefont {B.}~\bibnamefont {Roy}},\ }\bibfield
  {title} {\bibinfo {title} {Topolectric circuit: Theory and construction},\
  }\href@noop {} {\bibfield  {journal} {\bibinfo  {journal} {Phys. Rev. Res.}\
  }\textbf {\bibinfo {volume} {3}},\ \bibinfo {pages} {023056} (\bibinfo {year}
  {2021})}\BibitemShut {NoStop}%
\bibitem [{\citenamefont {amd S.~Imhof}\ \emph {et~al.}(2018)\citenamefont {amd
  S.~Imhof}, \citenamefont {Berger}, \citenamefont {Bayer}, \citenamefont
  {Brehm}, \citenamefont {Molenkamp}, \citenamefont {Kiessling},\ and\
  \citenamefont {Thomale}}]{topcircuit1}%
  \BibitemOpen
  \bibfield  {author} {\bibinfo {author} {\bibfnamefont {C.~H.~L.}\
  \bibnamefont {amd S.~Imhof}}, \bibinfo {author} {\bibfnamefont
  {C.}~\bibnamefont {Berger}}, \bibinfo {author} {\bibfnamefont
  {F.}~\bibnamefont {Bayer}}, \bibinfo {author} {\bibfnamefont
  {J.}~\bibnamefont {Brehm}}, \bibinfo {author} {\bibfnamefont {L.~W.}\
  \bibnamefont {Molenkamp}}, \bibinfo {author} {\bibfnamefont {T.}~\bibnamefont
  {Kiessling}},\ and\ \bibinfo {author} {\bibfnamefont {R.}~\bibnamefont
  {Thomale}},\ }\bibfield  {title} {\bibinfo {title} {Topological circuits},\
  }\href@noop {} {\bibfield  {journal} {\bibinfo  {journal} {Commun. Phys.}\
  }\textbf {\bibinfo {volume} {1}},\ \bibinfo {pages} {39} (\bibinfo {year}
  {2018})}\BibitemShut {NoStop}%
\bibitem [{\citenamefont {Johnson}\ and\ \citenamefont
  {Rasolofosaon}(1996)}]{nonlinelast}%
  \BibitemOpen
  \bibfield  {author} {\bibinfo {author} {\bibfnamefont {P.}~\bibnamefont
  {Johnson}}\ and\ \bibinfo {author} {\bibfnamefont {P.}~\bibnamefont
  {Rasolofosaon}},\ }\bibfield  {title} {\bibinfo {title} {Nonlinear elasticity
  and stress-induced anisotropy in rock},\ }\href@noop {} {\bibfield  {journal}
  {\bibinfo  {journal} {J. Geophys. Res.}\ }\textbf {\bibinfo {volume} {101}},\
  \bibinfo {pages} {3113} (\bibinfo {year} {1996})}\BibitemShut {NoStop}%
\bibitem [{\citenamefont {Senyuk}\ \emph {et~al.}(2021)\citenamefont {Senyuk},
  \citenamefont {Mozaffari}, \citenamefont {Crust}, \citenamefont {Zhang},
  \citenamefont {de~Pablo},\ and\ \citenamefont {Smalyukh}}]{smalyukh1}%
  \BibitemOpen
  \bibfield  {author} {\bibinfo {author} {\bibfnamefont {B.}~\bibnamefont
  {Senyuk}}, \bibinfo {author} {\bibfnamefont {A.}~\bibnamefont {Mozaffari}},
  \bibinfo {author} {\bibfnamefont {K.}~\bibnamefont {Crust}}, \bibinfo
  {author} {\bibfnamefont {R.}~\bibnamefont {Zhang}}, \bibinfo {author}
  {\bibfnamefont {J.~J.}\ \bibnamefont {de~Pablo}},\ and\ \bibinfo {author}
  {\bibfnamefont {I.~I.}\ \bibnamefont {Smalyukh}},\ }\bibfield  {title}
  {\bibinfo {title} {Transformation between elastic dipoles, quadrupoles,
  octupoles, and hexadecapoles driven by surfactant self-assembly in nematic
  emulsion},\ }\href@noop {} {\bibfield  {journal} {\bibinfo  {journal} {Sci.
  Adv.}\ }\textbf {\bibinfo {volume} {7}},\ \bibinfo {pages} {eabg0377}
  (\bibinfo {year} {2021})}\BibitemShut {NoStop}%
\bibitem [{\citenamefont {Senyuk}\ \emph {et~al.}(2019)\citenamefont {Senyuk},
  \citenamefont {Aplinc}, \citenamefont {Ravnik},\ and\ \citenamefont
  {Smalyukh}}]{smalyukh2}%
  \BibitemOpen
  \bibfield  {author} {\bibinfo {author} {\bibfnamefont {B.}~\bibnamefont
  {Senyuk}}, \bibinfo {author} {\bibfnamefont {J.}~\bibnamefont {Aplinc}},
  \bibinfo {author} {\bibfnamefont {M.}~\bibnamefont {Ravnik}},\ and\ \bibinfo
  {author} {\bibfnamefont {I.~I.}\ \bibnamefont {Smalyukh}},\ }\bibfield
  {title} {\bibinfo {title} {High-order elastic multipoles ads colloidal
  atoms},\ }\href@noop {} {\bibfield  {journal} {\bibinfo  {journal} {Nature
  Commun.}\ }\textbf {\bibinfo {volume} {10}},\ \bibinfo {pages} {1825}
  (\bibinfo {year} {2019})}\BibitemShut {NoStop}%
\bibitem [{\citenamefont {Carretta}\ \emph {et~al.}(2010)\citenamefont
  {Carretta}, \citenamefont {Santini}, \citenamefont {Caciuffo},\ and\
  \citenamefont {Amoretti}}]{carretta}%
  \BibitemOpen
  \bibfield  {author} {\bibinfo {author} {\bibfnamefont {S.}~\bibnamefont
  {Carretta}}, \bibinfo {author} {\bibfnamefont {P.}~\bibnamefont {Santini}},
  \bibinfo {author} {\bibfnamefont {R.}~\bibnamefont {Caciuffo}},\ and\
  \bibinfo {author} {\bibfnamefont {G.}~\bibnamefont {Amoretti}},\ }\bibfield
  {title} {\bibinfo {title} {Quadrupolar waves in uranium dioxide},\
  }\href@noop {} {\bibfield  {journal} {\bibinfo  {journal} {Phys. Rev. Lett.}\
  }\textbf {\bibinfo {volume} {105}},\ \bibinfo {pages} {167201} (\bibinfo
  {year} {2010})}\BibitemShut {NoStop}%
\bibitem [{\citenamefont {Strogatz}(2000)}]{Strogatz2000}%
  \BibitemOpen
  \bibfield  {author} {\bibinfo {author} {\bibfnamefont {S.~H.}\ \bibnamefont
  {Strogatz}},\ }\bibfield  {title} {\bibinfo {title} {From kuramoto to
  crawford: exploring the onset of synchronization in populations of coupled
  oscillators},\ }\href@noop {} {\bibfield  {journal} {\bibinfo  {journal}
  {Physica D}\ }\textbf {\bibinfo {volume} {143}},\ \bibinfo {pages} {1}
  (\bibinfo {year} {2000})}\BibitemShut {NoStop}%
\bibitem [{\citenamefont {Deco}\ \emph {et~al.}(2025)\citenamefont {Deco},
  \citenamefont {Perl}, \citenamefont {Greenstein}, \citenamefont {Chandaria},
  \citenamefont {Scholes},\ and\ \citenamefont {Kringelbach}}]{DecoQLbrain}%
  \BibitemOpen
  \bibfield  {author} {\bibinfo {author} {\bibfnamefont {G.}~\bibnamefont
  {Deco}}, \bibinfo {author} {\bibfnamefont {Y.}~\bibnamefont {Perl}}, \bibinfo
  {author} {\bibfnamefont {N.}~\bibnamefont {Greenstein}}, \bibinfo {author}
  {\bibfnamefont {S.}~\bibnamefont {Chandaria}}, \bibinfo {author}
  {\bibfnamefont {G.}~\bibnamefont {Scholes}},\ and\ \bibinfo {author}
  {\bibfnamefont {M.}~\bibnamefont {Kringelbach}},\ }\bibfield  {title}
  {\bibinfo {title} {Quantum-like dynamics in the human brain},\ }\href@noop {}
  {\bibfield  {journal} {\bibinfo  {journal} {xxxx}\ }\textbf {\bibinfo
  {volume} {xxx}},\ \bibinfo {pages} {bioRxiv:2025.2010} (\bibinfo {year}
  {2025})}\BibitemShut {NoStop}%
\bibitem [{\citenamefont {Boussard}\ \emph {et~al.}(2021)\citenamefont
  {Boussard}, \citenamefont {Fessel}, \citenamefont {Oettmeier}, \citenamefont
  {Briard}, \citenamefont {Döbereiner},\ and\ \citenamefont
  {Dussutour}}]{slime1}%
  \BibitemOpen
  \bibfield  {author} {\bibinfo {author} {\bibfnamefont {A.}~\bibnamefont
  {Boussard}}, \bibinfo {author} {\bibfnamefont {A.}~\bibnamefont {Fessel}},
  \bibinfo {author} {\bibfnamefont {C.}~\bibnamefont {Oettmeier}}, \bibinfo
  {author} {\bibfnamefont {L.}~\bibnamefont {Briard}}, \bibinfo {author}
  {\bibfnamefont {H.-G.}\ \bibnamefont {Döbereiner}},\ and\ \bibinfo {author}
  {\bibfnamefont {A.}~\bibnamefont {Dussutour}},\ }\bibfield  {title} {\bibinfo
  {title} {Adaptive behaviour and learnig in slime moulds: the role of
  oscillations},\ }\href@noop {} {\bibfield  {journal} {\bibinfo  {journal}
  {Phil. Trans. B}\ }\textbf {\bibinfo {volume} {376}},\ \bibinfo {pages}
  {20190757} (\bibinfo {year} {2021})}\BibitemShut {NoStop}%
\bibitem [{\citenamefont {Wootters}\ and\ \citenamefont
  {Zurek}(1982)}]{nocloning}%
  \BibitemOpen
  \bibfield  {author} {\bibinfo {author} {\bibfnamefont {W.}~\bibnamefont
  {Wootters}}\ and\ \bibinfo {author} {\bibfnamefont {W.~H.}\ \bibnamefont
  {Zurek}},\ }\bibfield  {title} {\bibinfo {title} {A single quantum cannot be
  cloned},\ }\href@noop {} {\bibfield  {journal} {\bibinfo  {journal} {Nature}\
  }\textbf {\bibinfo {volume} {299}},\ \bibinfo {pages} {802} (\bibinfo {year}
  {1982})}\BibitemShut {NoStop}%
\bibitem [{\citenamefont {Forbes}\ \emph {et~al.}(2025)\citenamefont {Forbes},
  \citenamefont {Nothlawala},\ and\ \citenamefont {Vallés}}]{ForbesNatPhoton}%
  \BibitemOpen
  \bibfield  {author} {\bibinfo {author} {\bibfnamefont {A.}~\bibnamefont
  {Forbes}}, \bibinfo {author} {\bibfnamefont {F.}~\bibnamefont {Nothlawala}},\
  and\ \bibinfo {author} {\bibfnamefont {A.}~\bibnamefont {Vallés}},\
  }\bibfield  {title} {\bibinfo {title} {Progress in quantum structured
  light},\ }\href@noop {} {\bibfield  {journal} {\bibinfo  {journal} {Nature
  Photonics}\ }\textbf {\bibinfo {volume} {19}},\ \bibinfo {pages} {1291}
  (\bibinfo {year} {2025})}\BibitemShut {NoStop}%
\bibitem [{\citenamefont {Shen}\ and\ \citenamefont
  {Rosales-Guzmán}(2022)}]{nonseplight}%
  \BibitemOpen
  \bibfield  {author} {\bibinfo {author} {\bibfnamefont {Y.}~\bibnamefont
  {Shen}}\ and\ \bibinfo {author} {\bibfnamefont {C.}~\bibnamefont
  {Rosales-Guzmán}},\ }\bibfield  {title} {\bibinfo {title} {Nonseparable
  states of light: From quantum to classical},\ }\href@noop {} {\bibfield
  {journal} {\bibinfo  {journal} {Laser Photonics Rev.}\ }\textbf {\bibinfo
  {volume} {16}},\ \bibinfo {pages} {2100533} (\bibinfo {year}
  {2022})}\BibitemShut {NoStop}%
\bibitem [{\citenamefont {Forbes}\ \emph {et~al.}(2019)\citenamefont {Forbes},
  \citenamefont {Aiello},\ and\ \citenamefont {Ndagano}}]{Forbes2019}%
  \BibitemOpen
  \bibfield  {author} {\bibinfo {author} {\bibfnamefont {A.}~\bibnamefont
  {Forbes}}, \bibinfo {author} {\bibfnamefont {A.}~\bibnamefont {Aiello}},\
  and\ \bibinfo {author} {\bibfnamefont {B.}~\bibnamefont {Ndagano}},\
  }\bibinfo {title} {Progess in optics}\ (\bibinfo  {publisher} {Elsevier},\
  \bibinfo {address} {Amsterdam, Netherlands},\ \bibinfo {year} {2019})\ Chap.\
  \bibinfo {chapter} {3: Classically entangled light}, pp.\ \bibinfo {pages}
  {99--153}\BibitemShut {NoStop}%
\bibitem [{\citenamefont {Barnett}\ \emph {et~al.}(2017)\citenamefont
  {Barnett}, \citenamefont {Babiker},\ and\ \citenamefont
  {Padgett}}]{Barnett-twisted}%
  \BibitemOpen
  \bibfield  {author} {\bibinfo {author} {\bibfnamefont {S.~M.}\ \bibnamefont
  {Barnett}}, \bibinfo {author} {\bibfnamefont {M.}~\bibnamefont {Babiker}},\
  and\ \bibinfo {author} {\bibfnamefont {M.~J.}\ \bibnamefont {Padgett}},\
  }\bibfield  {title} {\bibinfo {title} {Optical orbital angular momentum},\
  }\href@noop {} {\bibfield  {journal} {\bibinfo  {journal} {Phil. Trans. R.
  Soc.}\ }\textbf {\bibinfo {volume} {375}},\ \bibinfo {pages} {20150444}
  (\bibinfo {year} {2017})}\BibitemShut {NoStop}%
\bibitem [{\citenamefont {Spreeuw}(1998)}]{Spreeuw1998}%
  \BibitemOpen
  \bibfield  {author} {\bibinfo {author} {\bibfnamefont {R.~J.~C.}\
  \bibnamefont {Spreeuw}},\ }\bibfield  {title} {\bibinfo {title} {A classical
  analogy of entanglement},\ }\href@noop {} {\bibfield  {journal} {\bibinfo
  {journal} {Found. Phys.}\ }\textbf {\bibinfo {volume} {28}},\ \bibinfo
  {pages} {361} (\bibinfo {year} {1998})}\BibitemShut {NoStop}%
\bibitem [{\citenamefont {McLaren}\ \emph {et~al.}(2015)\citenamefont
  {McLaren}, \citenamefont {Konrad},\ and\ \citenamefont
  {Forbes}}]{Forbes2015}%
  \BibitemOpen
  \bibfield  {author} {\bibinfo {author} {\bibfnamefont {M.}~\bibnamefont
  {McLaren}}, \bibinfo {author} {\bibfnamefont {T.}~\bibnamefont {Konrad}},\
  and\ \bibinfo {author} {\bibfnamefont {A.}~\bibnamefont {Forbes}},\
  }\bibfield  {title} {\bibinfo {title} {Measuring the nonseparability of
  vector vortex beams},\ }\href@noop {} {\bibfield  {journal} {\bibinfo
  {journal} {Phys. Rev. A}\ }\textbf {\bibinfo {volume} {92}},\ \bibinfo
  {pages} {023833} (\bibinfo {year} {2015})}\BibitemShut {NoStop}%
\bibitem [{\citenamefont {Konrad}\ and\ \citenamefont
  {Forbes}(2019)}]{KonradForbes}%
  \BibitemOpen
  \bibfield  {author} {\bibinfo {author} {\bibfnamefont {T.}~\bibnamefont
  {Konrad}}\ and\ \bibinfo {author} {\bibfnamefont {A.}~\bibnamefont
  {Forbes}},\ }\bibfield  {title} {\bibinfo {title} {Quantum mechanics and
  classical light},\ }\href@noop {} {\bibfield  {journal} {\bibinfo  {journal}
  {Contemp. Phys.}\ }\textbf {\bibinfo {volume} {60}},\ \bibinfo {pages} {1}
  (\bibinfo {year} {2019})}\BibitemShut {NoStop}%
\bibitem [{\citenamefont {Brunner}\ \emph {et~al.}(2014)\citenamefont
  {Brunner}, \citenamefont {Calvalcanti}, \citenamefont {Pironio},
  \citenamefont {Scarani},\ and\ \citenamefont {Wehner}}]{Brunner}%
  \BibitemOpen
  \bibfield  {author} {\bibinfo {author} {\bibfnamefont {N.}~\bibnamefont
  {Brunner}}, \bibinfo {author} {\bibfnamefont {D.}~\bibnamefont
  {Calvalcanti}}, \bibinfo {author} {\bibfnamefont {S.}~\bibnamefont
  {Pironio}}, \bibinfo {author} {\bibfnamefont {V.}~\bibnamefont {Scarani}},\
  and\ \bibinfo {author} {\bibfnamefont {S.}~\bibnamefont {Wehner}},\
  }\bibfield  {title} {\bibinfo {title} {Bell nonlocality},\ }\href@noop {}
  {\bibfield  {journal} {\bibinfo  {journal} {Rev. Mod. Phys.}\ }\textbf
  {\bibinfo {volume} {86}},\ \bibinfo {pages} {419} (\bibinfo {year}
  {2014})}\BibitemShut {NoStop}%
\bibitem [{\citenamefont {Ndagano}\ \emph {et~al.}(2017)\citenamefont
  {Ndagano}, \citenamefont {Perez-Garcia}, \citenamefont {Roux}, \citenamefont
  {McLaren}, \citenamefont {Rosales-Guzman}, \citenamefont {Zhang},
  \citenamefont {Mouane}, \citenamefont {Hernandez-Aranda}, \citenamefont
  {Konrad},\ and\ \citenamefont {Forbes}}]{ForbesQChannels}%
  \BibitemOpen
  \bibfield  {author} {\bibinfo {author} {\bibfnamefont {B.}~\bibnamefont
  {Ndagano}}, \bibinfo {author} {\bibfnamefont {B.}~\bibnamefont
  {Perez-Garcia}}, \bibinfo {author} {\bibfnamefont {F.~S.}\ \bibnamefont
  {Roux}}, \bibinfo {author} {\bibfnamefont {M.}~\bibnamefont {McLaren}},
  \bibinfo {author} {\bibfnamefont {C.}~\bibnamefont {Rosales-Guzman}},
  \bibinfo {author} {\bibfnamefont {Y.}~\bibnamefont {Zhang}}, \bibinfo
  {author} {\bibfnamefont {O.}~\bibnamefont {Mouane}}, \bibinfo {author}
  {\bibfnamefont {R.~I.}\ \bibnamefont {Hernandez-Aranda}}, \bibinfo {author}
  {\bibfnamefont {T.}~\bibnamefont {Konrad}},\ and\ \bibinfo {author}
  {\bibfnamefont {A.}~\bibnamefont {Forbes}},\ }\bibfield  {title} {\bibinfo
  {title} {Characterizing quantum channels with non-separable states of
  classical light},\ }\href@noop {} {\bibfield  {journal} {\bibinfo  {journal}
  {Nature Physics}\ }\textbf {\bibinfo {volume} {13}},\ \bibinfo {pages} {397}
  (\bibinfo {year} {2017})}\BibitemShut {NoStop}%
\bibitem [{\citenamefont {Perez-Garcia}\ \emph {et~al.}(2015)\citenamefont
  {Perez-Garcia}, \citenamefont {Francis}, \citenamefont {McLaren},
  \citenamefont {Hernandez-Aranda}, \citenamefont {Forbes},\ and\ \citenamefont
  {Konrad}}]{ForbesQcomp}%
  \BibitemOpen
  \bibfield  {author} {\bibinfo {author} {\bibfnamefont {B.}~\bibnamefont
  {Perez-Garcia}}, \bibinfo {author} {\bibfnamefont {J.}~\bibnamefont
  {Francis}}, \bibinfo {author} {\bibfnamefont {M.}~\bibnamefont {McLaren}},
  \bibinfo {author} {\bibfnamefont {R.~I.}\ \bibnamefont {Hernandez-Aranda}},
  \bibinfo {author} {\bibfnamefont {A.}~\bibnamefont {Forbes}},\ and\ \bibinfo
  {author} {\bibfnamefont {T.}~\bibnamefont {Konrad}},\ }\bibfield  {title}
  {\bibinfo {title} {Quantum computation with classical light: The deutsch
  algorithm},\ }\href@noop {} {\bibfield  {journal} {\bibinfo  {journal} {Phys.
  Lett. A}\ }\textbf {\bibinfo {volume} {379}},\ \bibinfo {pages} {1675}
  (\bibinfo {year} {2015})}\BibitemShut {NoStop}%
\bibitem [{\citenamefont {Paneru}\ \emph {et~al.}(2020)\citenamefont {Paneru},
  \citenamefont {Cohen}, \citenamefont {Fickler}, \citenamefont {Boyd},\ and\
  \citenamefont {Karimi}}]{Paneru2020}%
  \BibitemOpen
  \bibfield  {author} {\bibinfo {author} {\bibfnamefont {D.}~\bibnamefont
  {Paneru}}, \bibinfo {author} {\bibfnamefont {E.}~\bibnamefont {Cohen}},
  \bibinfo {author} {\bibfnamefont {R.}~\bibnamefont {Fickler}}, \bibinfo
  {author} {\bibfnamefont {R.~W.}\ \bibnamefont {Boyd}},\ and\ \bibinfo
  {author} {\bibfnamefont {E.}~\bibnamefont {Karimi}},\ }\bibfield  {title}
  {\bibinfo {title} {Entanglement: quantum or classical?},\ }\href@noop {}
  {\bibfield  {journal} {\bibinfo  {journal} {Rep. Progr. Phys.}\ }\textbf
  {\bibinfo {volume} {83}},\ \bibinfo {pages} {064001} (\bibinfo {year}
  {2020})}\BibitemShut {NoStop}%
\bibitem [{\citenamefont {Strogatz}\ and\ \citenamefont
  {Stewart}(1993)}]{Strogatz1993}%
  \BibitemOpen
  \bibfield  {author} {\bibinfo {author} {\bibfnamefont {S.~H.}\ \bibnamefont
  {Strogatz}}\ and\ \bibinfo {author} {\bibfnamefont {I.}~\bibnamefont
  {Stewart}},\ }\bibfield  {title} {\bibinfo {title} {Coupled oscillators and
  biological synchronization},\ }\href@noop {} {\bibfield  {journal} {\bibinfo
  {journal} {Sci. Am.}\ }\textbf {\bibinfo {volume} {269}},\ \bibinfo {pages}
  {102} (\bibinfo {year} {1993})}\BibitemShut {NoStop}%
\bibitem [{\citenamefont {Strogatz}(2003)}]{StrogatzBook}%
  \BibitemOpen
  \bibfield  {author} {\bibinfo {author} {\bibfnamefont {S.}~\bibnamefont
  {Strogatz}},\ }\href@noop {} {\emph {\bibinfo {title} {Sync: How order
  emerges from chaos in the Universe, Nature, and Daily Life}}}\ (\bibinfo
  {publisher} {Hyperion, New York},\ \bibinfo {year} {2003})\BibitemShut
  {NoStop}%
\bibitem [{\citenamefont {Acebrón}\ \emph {et~al.}(2005)\citenamefont
  {Acebrón}, \citenamefont {Bonilla}, \citenamefont {Pérez-Vicente},
  \citenamefont {Ritort},\ and\ \citenamefont {Spigler}}]{Acebron2005}%
  \BibitemOpen
  \bibfield  {author} {\bibinfo {author} {\bibfnamefont {J.}~\bibnamefont
  {Acebrón}}, \bibinfo {author} {\bibfnamefont {L.}~\bibnamefont {Bonilla}},
  \bibinfo {author} {\bibfnamefont {C.}~\bibnamefont {Pérez-Vicente}},
  \bibinfo {author} {\bibfnamefont {F.}~\bibnamefont {Ritort}},\ and\ \bibinfo
  {author} {\bibfnamefont {R.}~\bibnamefont {Spigler}},\ }\bibfield  {title}
  {\bibinfo {title} {The kuramoto model: a simple paradigm for synchronization
  phenomena},\ }\href@noop {} {\bibfield  {journal} {\bibinfo  {journal} {Rev.
  Mod. Phys.}\ }\textbf {\bibinfo {volume} {77}},\ \bibinfo {pages} {137}
  (\bibinfo {year} {2005})}\BibitemShut {NoStop}%
\bibitem [{\citenamefont {daFonseca}\ and\ \citenamefont
  {Abud}(2018)}]{daFonseca2018}%
  \BibitemOpen
  \bibfield  {author} {\bibinfo {author} {\bibfnamefont {J.~D.}\ \bibnamefont
  {daFonseca}}\ and\ \bibinfo {author} {\bibfnamefont {C.~V.}\ \bibnamefont
  {Abud}},\ }\bibfield  {title} {\bibinfo {title} {The kuramoto model
  revisited},\ }\href@noop {} {\bibfield  {journal} {\bibinfo  {journal} {J.
  Stat. Mech.}\ }\textbf {\bibinfo {volume} {2018}},\ \bibinfo {pages} {103204}
  (\bibinfo {year} {2018})}\BibitemShut {NoStop}%
\bibitem [{\citenamefont {Rodrigues}\ \emph {et~al.}(2016)\citenamefont
  {Rodrigues}, \citenamefont {Peron}, \citenamefont {Ji},\ and\ \citenamefont
  {Kurths}}]{Rodrigues2016}%
  \BibitemOpen
  \bibfield  {author} {\bibinfo {author} {\bibfnamefont {F.~A.}\ \bibnamefont
  {Rodrigues}}, \bibinfo {author} {\bibfnamefont {T.~K.~D.}\ \bibnamefont
  {Peron}}, \bibinfo {author} {\bibfnamefont {P.}~\bibnamefont {Ji}},\ and\
  \bibinfo {author} {\bibfnamefont {J.}~\bibnamefont {Kurths}},\ }\bibfield
  {title} {\bibinfo {title} {The kuramoto model in complex networks},\
  }\href@noop {} {\bibfield  {journal} {\bibinfo  {journal} {Phys. Rep.}\
  }\textbf {\bibinfo {volume} {610}},\ \bibinfo {pages} {1} (\bibinfo {year}
  {2016})}\BibitemShut {NoStop}%
\bibitem [{\citenamefont {Scholes}(2022)}]{ScholesAbsSync}%
  \BibitemOpen
  \bibfield  {author} {\bibinfo {author} {\bibfnamefont {G.~D.}\ \bibnamefont
  {Scholes}},\ }\bibfield  {title} {\bibinfo {title} {The kuramoto-lohe model
  and collective absorption of a photon},\ }\href@noop {} {\bibfield  {journal}
  {\bibinfo  {journal} {Proc. R. Soc. A}\ }\textbf {\bibinfo {volume} {478}},\
  \bibinfo {pages} {20220377} (\bibinfo {year} {2022})}\BibitemShut {NoStop}%
\bibitem [{\citenamefont {Skardal}\ and\ \citenamefont
  {Arenas}(2020)}]{Kuramotohigher1}%
  \BibitemOpen
  \bibfield  {author} {\bibinfo {author} {\bibfnamefont {P.~S.}\ \bibnamefont
  {Skardal}}\ and\ \bibinfo {author} {\bibfnamefont {A.}~\bibnamefont
  {Arenas}},\ }\bibfield  {title} {\bibinfo {title} {Higher order interactions
  in complex networks of phase oscillators promote abrupt synchronization
  switching},\ }\href@noop {} {\bibfield  {journal} {\bibinfo  {journal}
  {Commun. Phys.}\ }\textbf {\bibinfo {volume} {2}},\ \bibinfo {pages} {218}
  (\bibinfo {year} {2020})}\BibitemShut {NoStop}%
\bibitem [{\citenamefont {Battiston}\ \emph {et~al.}(2020)\citenamefont
  {Battiston}, \citenamefont {Cencetti}, \citenamefont {Iacopini},
  \citenamefont {Latora}, \citenamefont {Lucas}, \citenamefont {Patania},
  \citenamefont {Young},\ and\ \citenamefont {Petri}}]{Kuramotohigher2}%
  \BibitemOpen
  \bibfield  {author} {\bibinfo {author} {\bibfnamefont {F.}~\bibnamefont
  {Battiston}}, \bibinfo {author} {\bibfnamefont {G.}~\bibnamefont {Cencetti}},
  \bibinfo {author} {\bibfnamefont {I.}~\bibnamefont {Iacopini}}, \bibinfo
  {author} {\bibfnamefont {V.}~\bibnamefont {Latora}}, \bibinfo {author}
  {\bibfnamefont {M.}~\bibnamefont {Lucas}}, \bibinfo {author} {\bibfnamefont
  {A.}~\bibnamefont {Patania}}, \bibinfo {author} {\bibfnamefont {J.-G.}\
  \bibnamefont {Young}},\ and\ \bibinfo {author} {\bibfnamefont
  {G.}~\bibnamefont {Petri}},\ }\bibfield  {title} {\bibinfo {title} {Networks
  beyond pairwise interactions: Structure and dynamics},\ }\href@noop {}
  {\bibfield  {journal} {\bibinfo  {journal} {Phys. Rep.}\ }\textbf {\bibinfo
  {volume} {874}},\ \bibinfo {pages} {1} (\bibinfo {year} {2020})}\BibitemShut
  {NoStop}%
\bibitem [{\citenamefont {Hill}\ and\ \citenamefont
  {Wootters}(1997)}]{Concurrence}%
  \BibitemOpen
  \bibfield  {author} {\bibinfo {author} {\bibfnamefont {S.~A.}\ \bibnamefont
  {Hill}}\ and\ \bibinfo {author} {\bibfnamefont {W.~K.}\ \bibnamefont
  {Wootters}},\ }\bibfield  {title} {\bibinfo {title} {Entanglement of a pair
  of quantum bits},\ }\href@noop {} {\bibfield  {journal} {\bibinfo  {journal}
  {Phys. Rev. Lett.}\ }\textbf {\bibinfo {volume} {78}},\ \bibinfo {pages}
  {5022} (\bibinfo {year} {1997})}\BibitemShut {NoStop}%
\bibitem [{\citenamefont {Stanley}(2012)}]{Stanley1}%
  \BibitemOpen
  \bibfield  {author} {\bibinfo {author} {\bibfnamefont {R.~P.}\ \bibnamefont
  {Stanley}},\ }\href@noop {} {\emph {\bibinfo {title} {Enumerative
  Combinatorics, Volume 1, 2nd Edition}}}\ (\bibinfo  {publisher} {Cambridge
  University Press, Cambridge},\ \bibinfo {year} {2012})\BibitemShut {NoStop}%
\bibitem [{\citenamefont {Scholes}(2023)}]{ScholesEntropy}%
  \BibitemOpen
  \bibfield  {author} {\bibinfo {author} {\bibfnamefont {G.~D.}\ \bibnamefont
  {Scholes}},\ }\bibfield  {title} {\bibinfo {title} {Large coherent states
  formed from disordered k-regular random graphs},\ }\href@noop {} {\bibfield
  {journal} {\bibinfo  {journal} {Entropy}\ }\textbf {\bibinfo {volume} {25}},\
  \bibinfo {pages} {1519} (\bibinfo {year} {2023})}\BibitemShut {NoStop}%
\bibitem [{\citenamefont {Barab\'asi}(2016)}]{Barabasi}%
  \BibitemOpen
  \bibfield  {author} {\bibinfo {author} {\bibfnamefont {A.-L.}\ \bibnamefont
  {Barab\'asi}},\ }\href@noop {} {\emph {\bibinfo {title} {Network Science}}}\
  (\bibinfo  {publisher} {Cambridge University Press, Cambridge},\ \bibinfo
  {year} {2016})\BibitemShut {NoStop}%
\bibitem [{\citenamefont {Traversa}\ and\ \citenamefont
  {Ventura}(2015)}]{DiVentura1}%
  \BibitemOpen
  \bibfield  {author} {\bibinfo {author} {\bibfnamefont {F.~L.}\ \bibnamefont
  {Traversa}}\ and\ \bibinfo {author} {\bibfnamefont {M.~D.}\ \bibnamefont
  {Ventura}},\ }\bibfield  {title} {\bibinfo {title} {Universal memcomputing
  machines},\ }\href@noop {} {\bibfield  {journal} {\bibinfo  {journal} {IEEE
  Trans. Neural Networks and Learning Systems}\ }\textbf {\bibinfo {volume}
  {26}},\ \bibinfo {pages} {2702} (\bibinfo {year} {2015})}\BibitemShut
  {NoStop}%
\bibitem [{\citenamefont {Traversa}\ \emph {et~al.}(2015)\citenamefont
  {Traversa}, \citenamefont {Ramella}, \citenamefont {Bonani},\ and\
  \citenamefont {Ventra}}]{DiVentura2}%
  \BibitemOpen
  \bibfield  {author} {\bibinfo {author} {\bibfnamefont {F.~L.}\ \bibnamefont
  {Traversa}}, \bibinfo {author} {\bibfnamefont {C.}~\bibnamefont {Ramella}},
  \bibinfo {author} {\bibfnamefont {F.}~\bibnamefont {Bonani}},\ and\ \bibinfo
  {author} {\bibfnamefont {M.~D.}\ \bibnamefont {Ventra}},\ }\bibfield  {title}
  {\bibinfo {title} {Memcomputing np-complete problems in polynomial time using
  polynomial resources and collective states},\ }\href@noop {} {\bibfield
  {journal} {\bibinfo  {journal} {Sci. Adv.}\ }\textbf {\bibinfo {volume}
  {1}},\ \bibinfo {pages} {e1500031} (\bibinfo {year} {2015})}\BibitemShut
  {NoStop}%
\bibitem [{\citenamefont {Csaba}\ and\ \citenamefont
  {Porod}(2020)}]{Csaba2020}%
  \BibitemOpen
  \bibfield  {author} {\bibinfo {author} {\bibfnamefont {G.}~\bibnamefont
  {Csaba}}\ and\ \bibinfo {author} {\bibfnamefont {W.}~\bibnamefont {Porod}},\
  }\bibfield  {title} {\bibinfo {title} {Coupled oscillators for computing: A
  review and perspective},\ }\href@noop {} {\bibfield  {journal} {\bibinfo
  {journal} {Appl. Phys. Rev.}\ }\textbf {\bibinfo {volume} {7}},\ \bibinfo
  {pages} {011302} (\bibinfo {year} {2020})}\BibitemShut {NoStop}%
\bibitem [{\citenamefont {Pothos}\ and\ \citenamefont
  {Busemeyer}(2022)}]{Pothos2022}%
  \BibitemOpen
  \bibfield  {author} {\bibinfo {author} {\bibfnamefont {E.~M.}\ \bibnamefont
  {Pothos}}\ and\ \bibinfo {author} {\bibfnamefont {J.~R.}\ \bibnamefont
  {Busemeyer}},\ }\bibfield  {title} {\bibinfo {title} {Quantum cognition},\
  }\href@noop {} {\bibfield  {journal} {\bibinfo  {journal} {Annu. Rev.
  Psychol.}\ }\textbf {\bibinfo {volume} {73}},\ \bibinfo {pages} {749}
  (\bibinfo {year} {2022})}\BibitemShut {NoStop}%
\bibitem [{\citenamefont {Wang}\ \emph {et~al.}(2014)\citenamefont {Wang},
  \citenamefont {Solloway}, \citenamefont {Shiffrin},\ and\ \citenamefont
  {Busemeyer}}]{Busemeyer2014}%
  \BibitemOpen
  \bibfield  {author} {\bibinfo {author} {\bibfnamefont {Z.}~\bibnamefont
  {Wang}}, \bibinfo {author} {\bibfnamefont {T.}~\bibnamefont {Solloway}},
  \bibinfo {author} {\bibfnamefont {R.}~\bibnamefont {Shiffrin}},\ and\
  \bibinfo {author} {\bibfnamefont {J.}~\bibnamefont {Busemeyer}},\ }\bibfield
  {title} {\bibinfo {title} {Context effects produced by question orders reveal
  quantum nature of human judgements},\ }\href@noop {} {\bibfield  {journal}
  {\bibinfo  {journal} {Proc. Natl. Acad. Sci. USA}\ }\textbf {\bibinfo
  {volume} {111}},\ \bibinfo {pages} {9431} (\bibinfo {year}
  {2014})}\BibitemShut {NoStop}%
\end{thebibliography}%

\end{document}